\documentclass[a4paper,fleqn,usenatbib]{mnras}

\usepackage{newtxtext,newtxmath}

\usepackage[T1]{fontenc}
\usepackage{ae,aecompl}

\usepackage{graphicx}	% Including figure files
\usepackage{amsmath}
\usepackage{amssymb}	% Extra maths symbols
\usepackage{array}
\usepackage{textcomp}

\newcommand{\Msol}{\mbox{$M_{\odot}$}}
\newcommand{\Lsol}{\mbox{$L_{\odot}$}}

\title[Mergers among MS and SB galaxies]{Early- and late-stage mergers among main sequence and starburst galaxies at $0.2{\leq}z{\leq}2$}

\author[A. Cibinel et al.]{A. Cibinel,$^{1}$\thanks{A.Cibinel@sussex.ac.uk}
E. Daddi$^{2}$,
M. T. Sargent$^{1}$,
E. Le Floc'h$^{2}$,
D. Liu$^{3,2}$,
F. Bournaud$^{2}$,
\newauthor
P. A. Oesch$^{4}$,
P. Amram$^{5}$,
A. Calabr\`o$^2$,
P.-A. Duc$^{6}$,
M. Pannella$^{7}$,
A. Puglisi$^{2,8}$,
\newauthor
V. Perret$^9$,
D. Elbaz$^2$, and
V. Kokorev$^{1}$
\\
$^{1}$Astronomy Centre, Department of Physics and Astronomy, University of Sussex, Brighton, BN1 9QH, UK\\
$^{2}$CEA, IRFU, DAp, AIM, Universit\'e Paris-Saclay, Universit\'e Paris Diderot, Sorbonne Paris Cit\'e, CNRS, F-91191 Gif-sur-Yvette, France \\
$^{3}$Max Planck Institute for Astronomy, K\"onigstuhl 17, D-69117 Heidelberg, Germany \\
$^{4}$Department of Astronomy, Universit\'e de Gen\`eve, Chemin des Maillettes 51,CH-1290 Versoix, Switzerland \\
$^{5}$Aix-Marseille   Universit\'e,   CNRS,   CNES,   LAM   (Laboratoire d'Astrophysique de Marseille), UMR 7326, F-13388 Marseille, France\\
$^{6}$ Observatoire astronomique de Strasbourg, Universit\'e de Strasbourg, CNRS,UMR 7550, F-67000 Strasbourg, France\\
$^{7}$Faculty of Physics, Ludwig-Maximilians University, Scheinerstr. 1, D-81679 Munich, Germany \\
$^8$Dipartimento di Fisica e Astronomia, Universit\`a di Padova, Vicolo dell'Osservatorio 2, I-35122 Padova, Italy \\
$^9$Center for Theoretical Astrophysics and Cosmology, Institute for Computational Science \& Physik Institut, \\
      \, University of Z\"{u}rich, 190 Winterthurestrasse, Z\"{u}rich CH-8057, Switzerland
}

\date{Accepted XXX. Received YYY; in original form ZZZ}

\pubyear{2017}

\begin{document}
\label{firstpage}
\pagerange{\pageref{firstpage}--\pageref{lastpage}}
\maketitle

% Abstract of the paper [250 words, 200 for letter]
\begin{abstract}
We investigate the fraction of close pairs and morphologically identified mergers on and above the star-forming main sequence (MS) at 0.2\,$\leq$\,$z$\,$\leq$2.0. 
The novelty of our work lies in the use of a non-parametric morphological classification performed on resolved stellar mass maps, reducing the contamination by non-interacting, high-redshift  clumpy galaxies. We find that the merger fraction rapidly rises to $\geq$70\% above the MS, implying that -- already at $z{\gtrsim}1$ -- starburst (SB) events ($\Delta_{\rm MS}$\,$\geq$\,0.6) are almost always associated with a major merger (1:1 to 1:6 mass ratio). The majority of interacting galaxies in the SB region are morphologically disturbed, late-stage mergers. Pair fractions show little dependence on MS offset and pairs are more prevalent than late-stage mergers only in the lower half of the MS. In our sample, major mergers on the MS occur with a roughly equal frequency of $\sim$5-10\% at all masses ${\gtrsim}10^{10}\Msol$. The MS major merger fraction roughly doubles between $z=0.2$ and $z=2$, with morphological mergers driving the overall increase at $z{\gtrsim}1$. The differential redshift evolution of interacting pairs and morphologically classified mergers on the MS can be reconciled by evolving observability time-scales for both pairs and morphological disturbances. 
The observed variation of the late-stage merger fraction with $\Delta_{\rm MS}$ follows the perturbative 2-Star Formation Mode model, where any MS galaxy can experience a continuum of different SFR enhancements. This points to an SB-merger connection not only for extreme events, but also more moderate bursts which merely scatter galaxies upward within the MS, rather than fully elevating them above it.
\end{abstract}


%%%%%%%%%%%%%%%%%%%%%%%%%%%%%%%%%%%%%%%%%%%%%
%%%%%%%%%%%%%%%%% BODY OF PAPER %%%%%%%%%%%%%%%%%%
%%%%%%%%%%%%%%%%%%%%%%%%%%%%%%%%%%%%%%%%%%%%%

%%%%%%%%%%%%%%%%%%%%%%%%%
%%%%      Section: Introduction      %%%%%%%
%%%%%%%%%%%%%%%%%%%%%%%%%
\section{Introduction}

Since at least $z$\,$\simeq$\,4, star-forming galaxies follow a tight, almost linear relation between their stellar mass and the star formation rate (SFR), the so-called `star formation main sequence' \citep[MS,][]{Brinchmann+04,Daddi+07,Elbaz+07,Elbaz+11,Noeske+07,Salim+07,Pannella+09,Santini+09,Karim+11,Whitaker+12,Steinhardt+14,Schreiber+15,Tomczak+16}.
The existence of such a sequence has been interpreted as evidence that, for the vast majority of galaxies, star formation proceeds in a quasi-steady, self-regulated fashion \citep[e.g.,][]{Noeske+07,Bouche+10,Leitner+12} and that stochastic star-formation enhancements causing galaxies to move above this sequence are rare or short-lived events \citep{Rodighiero+11,Sargent+12,Schreiber+15}

In the local Universe (z$\lesssim$0.3), the strongest starburst (SB) galaxies with star formation rates well above the MS -- e.g. luminous infrared galaxies (LIRGs) with $10^{11}\Lsol<L_{\rm IR}<10^{12}\Lsol$ or ultra-luminous infrared galaxies (ULIRGs) $L_{\rm IR}>$\,$10^{12}\Lsol$ -- are almost always triggered by a merger event \citep{Joseph_Wright_1985,Armus+87,Sanders_Mirabel_1996,Murphy+1996,Clements+96,Duc+97,Farrah+01,Veilleux+02,Arribas+04,Haan+11} and the merger fraction increases with increasing galaxy infrared (IR) luminosity \citep[e.g.][]{Scudder+12,Ellison+13,Larson+16}.
While major mergers are known to contribute to the high SFR tail distribution, a number of works suggest that minor mergers can also lead to SFR enhancements and offsets from the MS of factors of a few, comparable to those triggered in major mergers \citep{Mihos_Hernquist_94b,Scudder+12b,Kaviraj+14,Willet+15,Carpineti+15,Starkenburg+16,Martin+17}.
Despite being short-lived (typical depletion time scales for SB galaxies are of the order of $\lesssim$\,10$^{8}$yr, \citealt{Solomon+97,Solomon_VandenBout_2005,Daddi+10,Genzel+10,Saintonge+11}), these intense, merger-induced SBs can have a major impact on the galaxy gas reservoir and internal structure and are one of the possible mechanisms leading to star formation quenching and bulge formation \citep[]{Toomre_Toomre_1972,Gerhard_1981,Barnes_1988,Sanders+88,Mihos_Hernquist_96,Naab+99,DiMatteo+05,Springel+05,Hopkins+06,Hopkins+08}.

The degree to which this SB-merger connection persists at higher redshift is debated.
A different interpretation of the nature of off-MS galaxies in $z{\gtrsim}1$ far-IR selected samples is for example proposed in \citealt{Mancuso+16}. These author suggests that high specific star formation rate (sSFR) outliers correspond to young, SF galaxies caught while still assembling the bulk of their stellar mass, which for this reason display a higher SFR than expected given their mass. 
Although they might not be associated with the most extreme SFR enhancements, internal instabilities are also able to move high-$z$ galaxies toward the upper ridge of the MS to an sSFR close to that of SBs \citep{Zolotov+15,Tacchella+16}.
The exact definition of SBs in different works also contributes to discrepancies in the literature. Some of the early studies \citep[e.g.][]{Kartaltepe+10} adopted an absolute IR luminosity threshold to quantify the level of `starburstiness', as is typically done in local samples. However, due to the evolution of the average SFR of all galaxies with redshift (\citealt{Daddi+07,Elbaz+07,Karim+11,Whitaker+12}, see further references in \citealt{Speagle+14}), at $z\gtrsim1$ LIRGs and ULIRGs  dominate the MS \citep{Daddi+07,Sargent+12} and the cosmic SFR density \citep{LeFloch+05,Caputi+07,Magnelli_2009,Bethermin+11,Casey+12}. They are hence representative of the `normal' galaxy population at these redshifts and not of extreme star formation episodes.

Nevertheless, these early studies found that even at $z{\sim}2$ a significant fraction of ULIRGs -- many of which are MS galaxies -- shows morphological signatures of merging activity \citep[$\sim$40\%, ][]{Kartaltepe+10}.
The existence of morphologically disturbed galaxies on the MS may indicate that  SFR enhancements due to merger events are typically small (factors of about $\times$2--3), as suggested by a number of observational works \citep[e.g.][]{Jogee+09,Robaina+09,Wong+11,Xu+12,Kaviraj+13}.
%A number of observational works suggest small SFR enhancement  in mergers with respect to non-interacting galaxies up to $z{\simeq}2$ \citep[e.g.][]{Jogee+09,Robaina+09,Wong+11,Xu+12,Kaviraj+13}.
Parsec-scale numerical simulations of high-$z$ galaxies also show that the high gas fractions and disc turbulence in these galaxies drive significant central gas inflow even in isolated galaxies, therefore reducing the impact of mergers which induce either no or very short duration SBs \citep{Perret+14,Fensch+17}. 

%=============================%
%==                FIGURE 1                    ==%
%=============================%
\begin{figure}
\begin{center}
\includegraphics[width=0.49\textwidth]{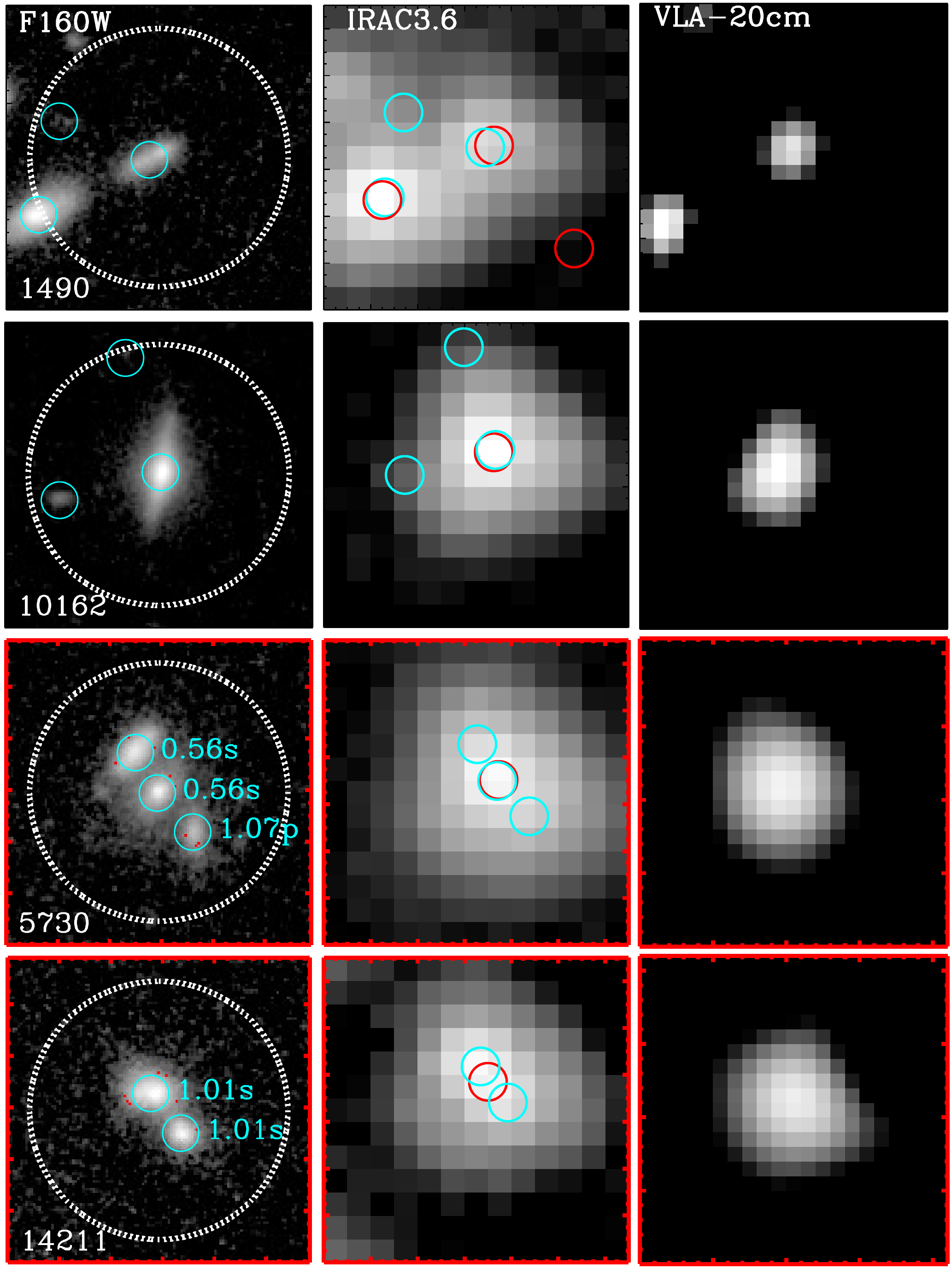}
\end{center}
\caption{\label{fig:GoodMatch} Examples of optical-IR matches. From left to right: $HST$ $H$-band image centred on the closest optical counterpart to the IRAC source, IRAC 3.6\,$\mu$m image and VLA 20\,cm radio image \citep{Owen18}. 
Cyan circles correspond to $H-$band detected galaxies in the 3D-$HST$ catalogue, red circles are galaxies in the GOODS-$Spitzer$ IRAC parent catalogue used in L18.\newline
Rows 1 \& 2: examples of robust optical-IR matches where the IRAC prior source for the FIR+mm flux extraction is associated with a single optical counterpart. Although more than one optical source is present within 3$\arcsec$ (dotted white circle) of the central IRAC source, the long wavelength flux is either deblended into multiple sources or dominated by the central galaxy.\newline
Rows 3 \& 4 (red frames): examples of uncertain associations, where the parent IRAC source is not clearly associated with a single optical counterpart. Row 3 shows an IRAC source with three counterparts, two at the same spectroscopic redshift (``s" appended to redshift value next to cyan circle) and a third with an incompatible photometric redshift (labelled ``p"). The bottom row shows an example of a genuine close spectroscopic pair.}
\end{figure}
%=============================%
A different approach for defining SBs, which incorporates cosmic SFR evolution, is based on the offset of galaxies from the locus of the MS, often parametrized as the logarithmic difference $\Delta_{\rm MS}=\log({\rm SFR})-\log({\rm SFR}_{\rm MS})$ between a galaxy SFR and the characteristic SFR on the MS at given stellar mass (SFR$_{\rm MS}$). Studies investigating merger fractions among high-$\Delta_{\rm MS}$ outliers to the MS suggest that, also at higher redshift, SB activity is often associated with ongoing mergers or interactions  \citep{Kartaltepe+12,Hung+13}.

The quantification of the role of mergers in MS and SB galaxies out to high redshift is however a challenging task. 
Together with observational limitations affecting galaxy classifications (e.g., surface brightness dimming, morphological $k$-correction), galaxies become intrinsically clumpier as redshift increases, resulting in asymmetric and disturbed appearances also in `normal' disc galaxies that are not undergoing an interaction \citep{Cowie+95,Papovich+05,Genzel+06,Elmegreen+07,Law+07,Bournaud+07,Bournaud+08,vanStarkenburg+08,Swinbank+10,Forster_Schreiber+09,Forster_Schreiber+11,Guo+12,Tacconi+13}.
In all aforementioned morphological merger studies, the distinction between merging galaxies and non-interacting systems is based on either a visual or a parametric classification of near-IR, single-band photometry, probing the rest-frame ultraviolet (UV)-to-optical emission, which is dominated by giant clumps.
 
In this paper, we re-examine the extent to which SB galaxies are directly linked with mergers, and in general how merging galaxies are distributed throughout the MS, by using a novel classification technique which instead relies on morphological information derived from the resolved mass maps that we developed in \citet[][Paper I hereafter]{Cibinel+15}. 
By using spatially resolved spectral energy distribution (SED) information to reconstruct the intrinsic stellar mass distribution of galaxies, this method is optimised for reducing the contamination by clumpy discs in high-$z$ merger samples, which have complex UV-optical morphologies but smooth mass distributions. 
We also combine these morphologically selected mergers with galaxies in close pair systems to study how the different merger stages affect the galaxy position on the MS.
For this study, we make use of the latest measurements of deep, far-infrared (FIR) to sub-millimetre photometry in the Great Observatories Origins Deep Survey North area \citep[GOODS-N,][]{Giavalisco+04}  derived in \citet{Liu+18}.  Exploiting an improved deblending algorithm, these refined IR measurements significantly reduce source confusion and artificial flux boosting and provide reliable measurements of integrated SFR also for dust-obscured SBs, for which UV or optical tracers result in biased estimates.

The paper is organized as follows: We present the galaxy sample and available data in Section~\ref{sec:Data}. Sections~\ref{sec:mergerClass} and \ref{sec:Corrections} describe the merger identification techniques used in this study and how we calculate the merger fraction. Our results are presented in  Section~\ref{sec:Results} and discussed in the context of other literature findings in Section~\ref{sec:Discussion}. 
A summary is provided in Section~\ref{sec:Summary}. 
We use magnitudes in the AB system \citep{Oke_1974} and a Chabrier initial mass function \citep[]{Chabrier_2003}.
When necessary, measurements for comparison samples are converted accordingly. 
We adopt a Planck 2015 cosmology ($\Omega_m$\,=\,0.31, $\Omega_\Lambda$\,+\,$\Omega_m$\,=\,1, and $H_0$\,=\,67.7\,km\,s$^{-1}$\,Mpc$^{-1}$, \citealt{Planck+16}).

Note that, throughout the paper, we use two different parametrizations of the redshift-dependent locus of the MS: the one derived from the literature compilation in \citet{Sargent+14} and the one derived from FIR detections or stacks in \citet{Schreiber+15}. 
The major difference between the two definitions is that the latter displays a ``bending" towards high stellar masses.
Unless otherwise stated, results will always be presented and discussed for both parametrizations.
We furthermore employ ``total" SFRs defined as SFR=SFR$_{\rm UV}$+SFR$_{\rm IR}$, where SFR$_{\rm UV}$ is the dust-unobscured contribution derived from the UV luminosity and SFR$_{\rm IR}$ represents the obscured SFR derived from light re-emitted in the 8--1000\,$\mu$m wavelength range.

Finally, in the literature major mergers are usually defined as those involving galaxies with mass ratios in the range 1:1 to 1:4, whereas minor mergers typically have ratios 1:4 to 1:10.
As explained later in the text, our study is restricted to mergers in the range 1:1 to 1:6. Our sample therefore includes some minor mergers, but the majority lies in the major merger regime and for simplicity we refer to mergers studied here as ``major mergers" from now on.

%%%%%%%%%%%%%%%%%%%%%%%%%%%%%%%%
%%%%%%%%             SEC.  DATA                %%%%%%%%%
%%%%%%%%%%%%%%%%%%%%%%%%%%%%%%%%

\section{Data} \label{sec:Data}
\subsection{Main FIR+mm sample}\label{sec:Selection}

For the analysis in this paper we consider a sample of far-IR and sub-millimiter (FIR+mm) detected galaxies in the GOODS-N area. The availability of FIR measurements is essential for reliably probing the star formation in the SB regime where strong dust obscuration can affect shorter wavelength SFR tracers.
We therefore selected our sample from the IR catalogue published in \citet[][L18 hereafter]{Liu+18}, which
consists of galaxies from the IRAC GOODS--Spitzer survey (PI: M. Dickinson), that also have both a MIPS 24\,$\mu$m and a Very Large Array (VLA) 20\,cm radio detection. 

Photometric data ranging from the mid-IR to the radio are available for this sample, including observations from the GOODS-$Spitzer$ Legacy Program (including the MIPS 24\,$\mu$m, PI: M. Dickison), $Herschel$-PACS 100 and 160\,$\mu$m (co-added PACS Evolutionary Probe, PEP, and  GOODS-$Herschel$ programs, \citealt{Lutz+11,Elbaz+11}), $Herschel$-SPIRE 250, 350, and 500\,$\mu$m \citep{Elbaz+11},  SCUBA-2  850\,$\mu$m \citep{Geach+17} and VLA 20\,cm \citep{Owen18}.
For all these galaxies, IR flux densities and SEDs were derived through an improved source deblending technique, which relies on the shorter wavelength SED properties of each primary target and its neighbours to define a list of prior sources that are likely to contribute to the confusion-prone maps at a given IR wavelength. 
The details of these measurements and of the calculation of IR-derived properties are provided in L18. 
%=============================%
%==                FIGURE 2                    ==%
%=============================%

\begin{figure*}
\begin{center}
\includegraphics[width=0.9\textwidth]{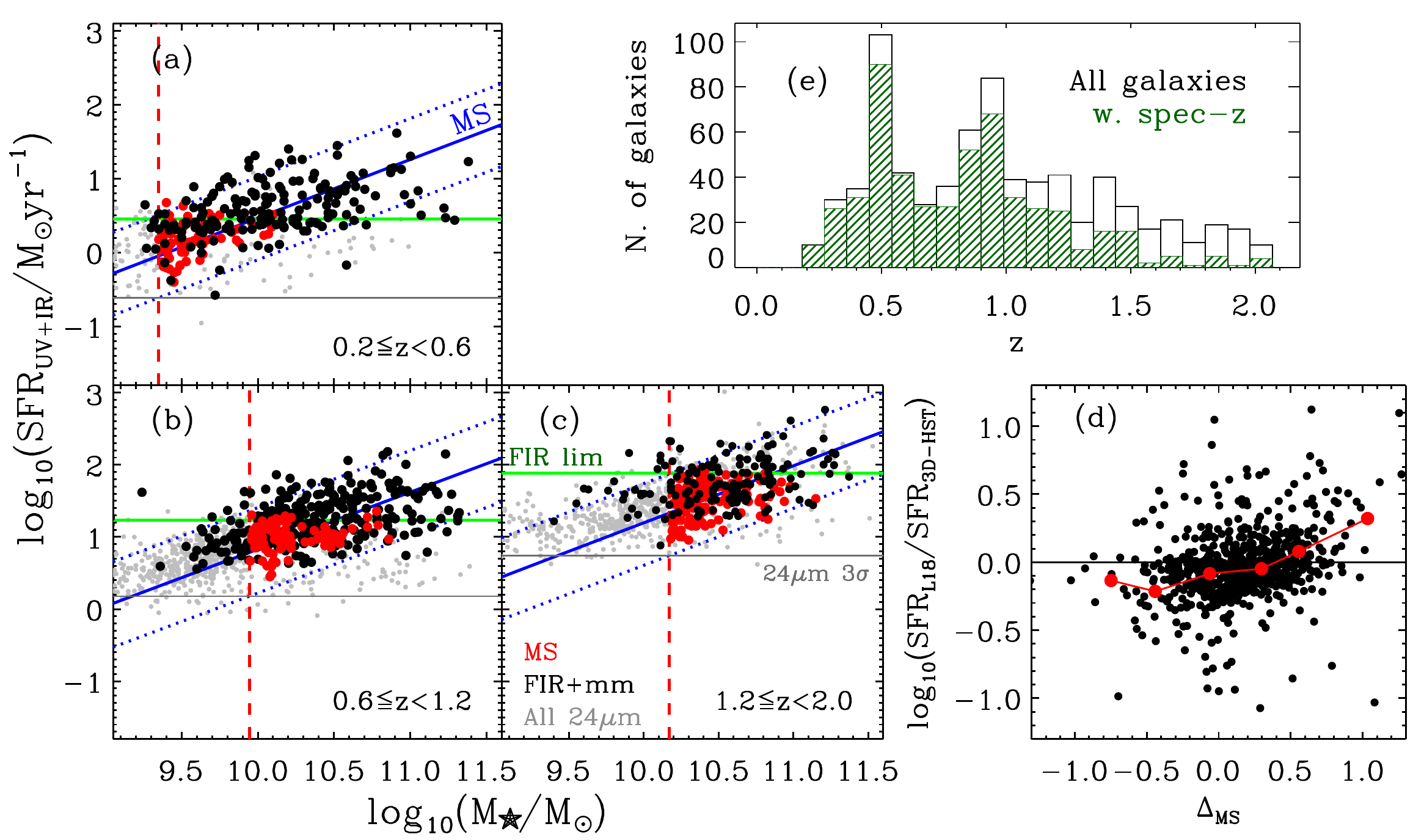}
\end{center}
\caption{\label{fig:Selection}  Properties of the FIR+mm and 3D-$HST$ MS samples. \emph{Panels (a)-(c):} Distribution in the SFR-$M_{\star}$ plane for our three redshift bins. Black points are FIR+mm detected galaxies in the L18 catalogue, red symbols the MS sample extracted from 3D-$HST$. For reference, all 24\,$\mu$m-detected (SNR$>$3) galaxies in 3D-$HST$ at the same redshift are plotted with small grey points. The horizontal green line shows the SFR limit corresponding to the FIR selection at the highest redshift in each bin, the horizontal grey line is the 24\,$\mu$m detection limit. The solid and dotted blue lines trace the MS \citep{Sargent+14} and 0.4\,dex offset locus at the mean redshift. 
The red dashed line indicates the minimum mass (completeness limit) of the combined FIR+mm and 3D-$HST$ MS sample. 
\emph{Panel (d):} Difference between the SFR derived from the full IR+mm SED fitting in L18 (SFR$_{\rm IR}$) and the 3D-$HST$ estimate (SFR$_{\rm3D-HST}$), as function of the distance $\Delta_{\rm MS}$ from the MS (referenced to SFR$_{\rm IR}$). Small black symbols show individual galaxies in the FIR+mm sample and big red points the median difference. For galaxies with $\left|\Delta_{\rm MS}\right|{\lesssim}$0.4\,dex the two estimates are consistent; significant deviations are observed above and below the MS.
\emph{Panel (e)}: redshift distribution of the combined FIR+mm and MS samples. The green histogram shows the redshift distribution of galaxies with a spectroscopic redshift.}
\end{figure*}
%=============================%

To build our FIR-detected sample we start by considering sources in the parent L18 catalogue which: (i) are within the inner GOODS-N area having a high level of completeness of 24\,$\mu$m prior sources (low 24\,$\mu$m instrumental noise, \emph{goodArea}=1 in Table 2 of L18) and, (ii) are FIR+mm-detected with a combined signal-to-noise ratio SNR$_{\rm FIR+mm}$$\ge$5\footnote{SNR$^2_{\rm FIR+mm}$=SNR$^2_{100\mu \mathrm{m}}$+SNR$^2_{160\mu \mathrm{m}}$+SNR$^2_{250\mu \mathrm{m}}$+SNR$^2_{350\mu \mathrm{m}}$
+SNR$^2_{500\mu \mathrm{m}}$+SNR$^2_{850\mu \mathrm{m}}$+SNR$^2_{1.6\mathrm{mm}}$,
as defined in L18.}. To ensure that reliable morphological parameters could be derived, we then selected a subset of galaxies from this FIR+mm-detected sample with: 
\begin{enumerate}
\item A spectroscopic (spec-$z$) or photometric redshift (photo-$z$) in the range 0.2$\le$z$\le$2.0. %(zq < 3 in L18)
About 80\% of the galaxies considered have a spec-$z$ from the redshift compilation described in L18 (of these, 60\% are 3D-$HST$ spec-$z$ -- i.e. non-grism -- measurements from \citealt{Skelton+14,Momcheva+16}). 
For the remainder of the sample we rely on the photo-$z$ as published in L18.
To keep evolutionary effects under control, we will present our findings for three separate redshift bins,  0.2\,$\leq$\,$z$\,$<$0.6, 0.6\,$\leq$\,$z$\,$<$1.2 and 1.2\,$\leq$\,$z$\,$\leq$2 (boundaries were selected such to divide the sample in bins with roughly equal number of galaxies, see Table~\ref{tab:SampleSize}). 
\item An observed $HST$/WFC3 $H_{160}$-band (`$H$-band' from now on) magnitude brighter than $H$\,$\le$\,24 and a Kron radius r$_{\rm kron}$\,$\ge$\,5$\times$FWHM(PSF), assigned by matching the L18 sample with the 3D-$HST$ catalogues (see next section). 
These criteria ensure that robust structural parameters can be measured on the resolved mass maps, as discussed in detail in Paper I\footnote{The $H$-band magnitude cut imposes a mass completeness limit and a SFR threshold. For star-forming galaxies these are, however, less stringent that the limits implied by the FIR detection discussed in Section~\ref{sec:SampleCompleteness}: the $H$-band selected sample is complete above masses $\log(M_{\star}/\Msol){\simeq}$10.0 and probes SFR$\simeq$10$\Msol$/yr at z=2.0. It is therefore not necessary to apply any statistical correction to the sample as a consequence of our $H$-band cut.}.
Only $\sim$2\% of the original FIR+mm-detected galaxies at  0.2$\le$z$\le$2.0 are rejected when applying these constraints (all in the two highest redshift bins).
While small in number, the exclusion of these galaxies could in principle bias our sample against late stage mergers dominated by a nuclear SB or extremely obscured merger-induced SB. However, we find that the fraction of these galaxies stays constant between the MS and SB regions and we therefore conclude that their exclusion will not affect our results.\end{enumerate}

A further 30 galaxies were discarded due to being too close to or outside the edges of the HST optical/NIR images or because of other artefacts.  We will refer to these FIR+mm-detected galaxies as the ``FIR+mm sample", from now on.

For all these galaxies, UV+IR SFR estimates have been published in L18. 
In particular, the dust-obscured component SFR$_{\rm IR}$ is derived from the SED fitting of the 2\,$\mu$m-to-radio, super-deblended photometry. The SED fitting included a stellar component \citep{Bruzual_Charlot_2003}, a mid-IR AGN torus template \citep{Mullaney+11}, a dust SED \citep{Magdis+12} and a power-law radio component as detailed in L18. SFR$_{\rm IR}$ are obtained from the pure dust component and corrected for the AGN contribution.
The unattenuated SFR$_{\rm UV}$ were obtained either by applying the \citet{Kennicutt_98} calibration to the rest-frame 1400\,$\AA$ fluxes or from the SFR$_{\rm UV}$/SFR$_{\rm IR}$  versus stellar mass correlation, for those galaxies that lack 1400\,$\AA$ fluxes.

\subsection{Optical counterparts of FIR+mm sources}\label{sec:PossibleBlends}

As mentioned above, an IRAC parent sample serves as basis for the identification 24\,$\mu$m+radio sources in the L18 catalogue from which we selected our FIR+mm sample.
To identify optical counterparts and assign stellar masses (as well as magnitudes), the L18 sample was then cross-matched with the 3D-HST catalogue \citep{Skelton+14} and the $K$-band sample of \citet{Pannella+15}.
Given the $\sim$1.5$\arcsec$ angular resolution of IRAC, blending of multiple galaxies within an IRAC resolution element, and as a consequence a FIR+mm source, is possible. Although it is not excluded that the IR emission could be associated with one of the galaxies only, an unambiguous match cannot always be established and such blends could bias the FIR-based SFR estimates and MS-offset calculation.

To keep these cases under control, we visually inspected all galaxies in the FIR+mm sample which are likely to suffer from such contamination/blending. We singled out as potentially problematic sources those which have, within a 3$\arcsec$ radius, more than one optical counterpart in the 3D-$HST$  catalogue with an IRAC-1 flux above the $1\sigma$ detection limit of $\sim$0.5\,$\mu$Jy  of the GOODS-N $Spitzer$ observations \citep{Elbaz+11}.
For these galaxies we checked the IRAC images and also the VLA 20\,cm images of \citet{Owen18} to determine whether the IRAC source could correspond to a blend of multiple galaxies resolved in the HST $H$-band image. We then treated galaxies according to the ``purity" of the optical--IR association, as follows.

Instances where the IRAC and radio emission are clearly dominated by a single optical galaxy, or where sources are not blended in the IRAC images, were validated as robust single matches and no further action was required (see examples in upper panels of Fig.~\ref{fig:GoodMatch}).
For cases where instead more than one galaxy may contribute to the IR fluxes ($\sim$10\% of the entire FIR+mm sample; see examples in bottom panels of Fig.~\ref{fig:GoodMatch}) we distinguished two categories:
\begin{itemize}
\item Genuine spectro-photometric associations of multiple galaxies at the same redshift ($\sim$1\% of the entire FIR+mm sample),  as defined in the merger classification described in more detail in the next section. In such cases, we consider all galaxies in the system as a single entity and assign to it a stellar mass given by the sum of the masses in the individual galaxies to avoid underestimation of the stellar component associated with the IR emission. 

\item Likely chance projections ($\sim$9\%). Such cases have the potential of artificially enhancing the SFR of 
non-interacting galaxies as consequence of the blending with the interloper. For these galaxies we keep the original mass-SFR association as published in L18 and we test how the results are affected when they are excluded from our sample. 
\end{itemize}

\subsection{Completeness and complementary MS sample}\label{sec:SampleCompleteness}

The detection constraint SNR$_{\rm FIR+mm}$$\ge$5 imposed to select the FIR+mm sample translates into a redshift-dependent minimum SFR. 
Consequently, while the FIR+mm sample is generally complete at high SFR enhancements, at a fixed stellar mass it becomes increasingly incomplete for galaxies on the MS as redshift increases. This is illustrated in Fig.~\ref{fig:Selection}. 

To ensure a sampling of the MS that is as homogeneous as possible across all redshift and mass bins, we added to the FIR+mm sample a set of MS galaxies extracted from the 3D-$HST$ survey (referred to as `3D-$HST$ MS sample'; cf. red symbols in Fig.~\ref{fig:Selection}).
Specifically, we include in this 3D-$HST$ MS sample galaxies that have a total SFR below the FIR+mm selection, satisfy the constraints $H$\,$\le$\,24 and r$_{\rm kron}$\,$\ge$\,5$\times$FWHM(PSF), and which lie no more than 0.4\,dex beneath the mean MS locus. 
By limiting our analysis to galaxies within this range of MS offsets we ensure that we remain complete down to stellar masses $\log(M_{\star}/\Msol){\simeq 10}$ at all redshifts $0.2{\leq}z{\leq}2.0$.

We note that care needs to be taken to ensure consistency of SFR measurements   when combining the FIR+mm and 3D-$HST$ MS samples.
As described in \citealt{Whitaker+14}, total SFRs for galaxies in the 3D-$HST$ MS sample were inferred from the IR and UV luminosity with an approach similar to \citet{Bell+05}, after
converting the Spitzer 24\,$\mu$m monochromatic flux density to a total (8-1000)\,$\mu$m $L_{\rm IR}$ luminosity and the rest-frame 2800\,$\AA$ flux to a total UV luminosity.
As shown in Fig.~\ref{fig:Selection}\emph{d}, 3D-$HST$ total SFR values agree very well with the L18 estimates ($<$0.1dex difference) for galaxies on the MS, i.e. within those parts of parameter space were we supplement the FIR+mm sample with 3D-$HST$ galaxies. For galaxies well above the MS, the SFR in the 3D-$HST$ catalogue is systematically smaller than that derived by L18, highlighting the importance of a full IR SED analysis for starbursting systems.  

The minimum stellar mass of galaxies included in the 3D-$HST$ MS samples at each redshift -- and therefore the minimum mass of the combined final FIR+mm and MS samples -- is the completeness limit above which the MS is well sampled by the 24\,$\mu$m detected galaxies (SNR$>$3).
These mass limits are $\log(M_{\star}/\Msol)$=9.4 at $0.2{\leq}z{<}0.6$, $\log(M_{\star}/\Msol)$=9.9 at $0.6{\leq}z{<}1.2$ and $\log(M_{\star}/\Msol)$=10.2 at $1.2{\leq}z{\leq}2.0$ (see dashed red lines in Fig.~\ref{fig:Selection}).

\subsection{Exclusion of AGN}

The SED fits underpinning the SFR estimates for the FIR+mm sample involve a mid-IR active galactic nucleus (AGN) component \citep{Mullaney+11}. As discussed above, SFRs were then derived based on the host galaxy dust component and are hence corrected for any AGN contribution to the IR emission.
Such a decomposition was however not performed on the 3D-$HST$ MS sample and the associated 24$\mu m$ flux densities and inferred SFR could suffer from AGN contamination.
Moreover, the presence of an AGN could also affect the UV to mid-IR SED fits used to derive the stellar population properties and stellar mass estimates.

For this reason, AGN candidates are removed from our combined FIR+mm and MS sample. We consider as AGNs those sources with either X-ray luminosity $L_{0.5-7 {\rm keV}}{\geq}3\times10^{42}$erg\,s$^{-1}$ or radio-excess indicative of radio-loud AGN activity as detailed in L18 (``Type.AGN=1").
To assign an X-ray luminosity to our galaxies, we matched our sample with the revised 2\,Ms $Chandra$ catalogue of point sources in the $Chandra$ Deep Field North \citep{Alexander+03,Xue+16}, using a 1.5${\arcsec}$ search radius around the optical counterparts identified by these authors.

After the exclusion of AGN candidates (58 galaxies), our final combined 3D-$HST$ MS and FIR+mm sample consists of 729 $0.2{\leq}z{\leq}$ 2.0 galaxies, as summarised in Table \ref{tab:SampleSize}.
%%===================%
%%==           TABLE 1       ==%
%%===================%
\begin{table}
	\centering
\caption{Number of galaxies and median stellar mass of our sample in each of the redshift bins considered in this study. The number of galaxies is provided for both the combined FIR+mm plus 3D-$HST$ MS sample and for the two separately.}
          \begin{tabular}{ccccc} 
		\hline\hline 
		 redshift     & N$_{\rm tot}$  & N$_{\rm FIR+mm}$  & N$_{\rm MS}$  & \textlangle$\log(M_{\star})$\textrangle  \\
	       \hline\hline 
	      all z                                    &   729  & 455 & 274 &  10.28     \\
	      0.2\,$\leq$\,$z$\,$<$0.6     &   219  & 146 &  73  & 9.86 \\
	      0.6\,$\leq$\,$z$\,$<$1.2     &   291  & 205 &  86  &  10.29 \\
	      1.2\,$\leq$\,$z$\,$\leq$2.0 &   219  & 104 &  115 & 10.43 \\                                   
                		\hline 		
	\end{tabular} \label{tab:SampleSize}
\end{table}
%%===================%

\subsection{Multiwavelength coverage, integrated stellar masses and resolved mass maps}

Multi-wavelength  $HST$/ACS and $HST$/WFC3 imaging from in the optical to NIR is available for all galaxies from GOODS, the Cosmic Assembly Near-infrared Deep Extragalactic Legacy Survey (CANDELS) \citep{Grogin+11,Koekemoer+11} and the 3D-$HST$ survey \citep{vanDokkum+11,Brammer+12}. In particular, to construct the resolved stellar mass maps for the morphological classification described below, we use the $B_{435}$, $V_{606}$, $i_{775}$, $i_{814}$, $z_{850}$,  $J_{125}$, $J_{140}$ and $H$-band images released by the 3D-$HST$ team, plus the original $Y_{105}$ from the CANDELS data.  
For 37\% of the galaxies in our sample UV imaging (WFC3 $F275W$ and $F336W$) is also available from the $Hubble$ Deep UV Imaging Survey \citep[HDUV,][]{Oesch+18}, which is covering the central $\sim$\,7.5$\arcmin$\,$\times$\,10$\arcmin$ region of GOODS-N.
%=============================%
%==                FIGURE 3                   ==%
%=============================%
\begin{figure*}
\begin{center}
\includegraphics[width=0.83\textwidth]{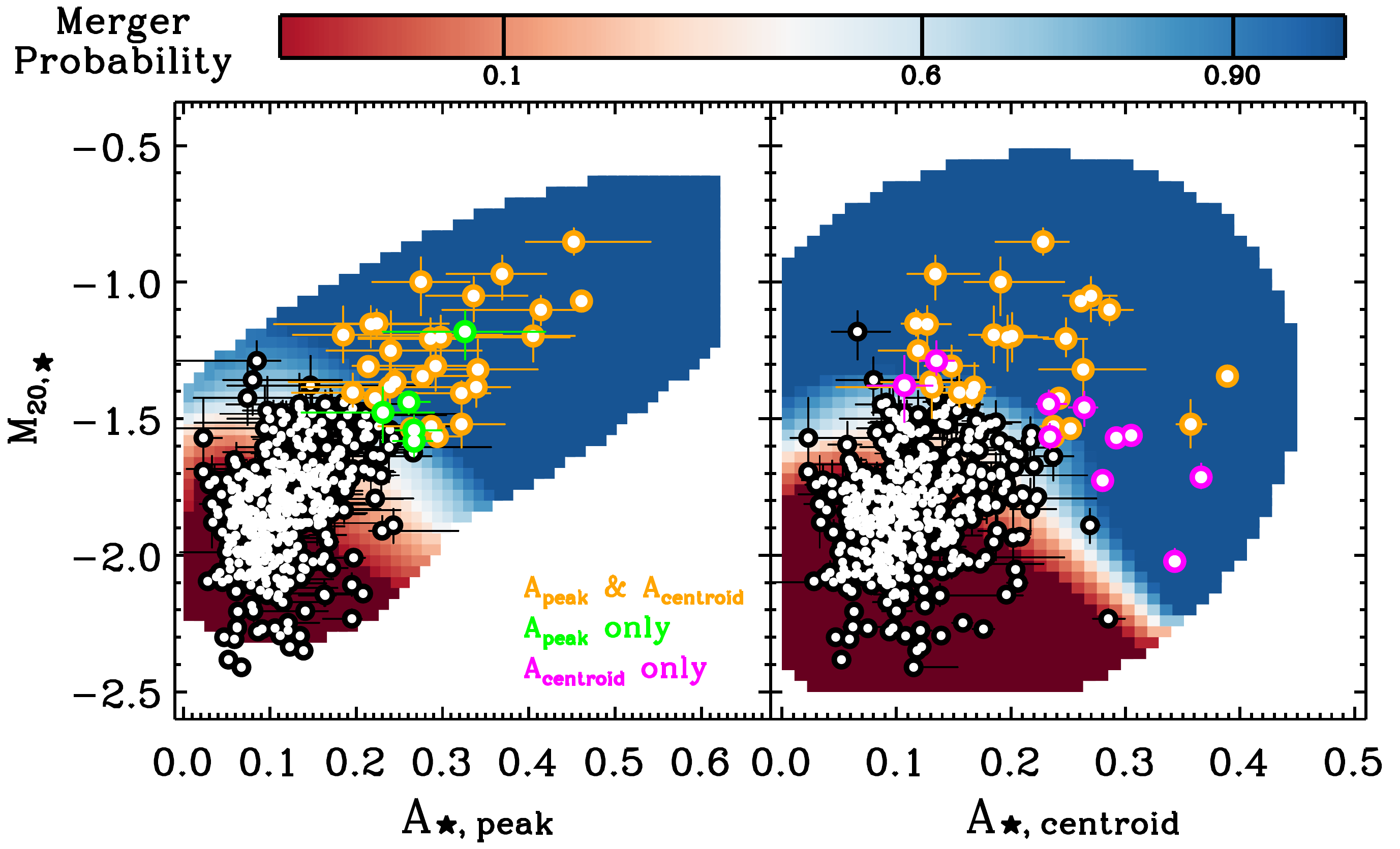}
\end{center}
\caption{\label{fig:Class} 
Distribution of the $M_{20, \star}$ and $A_{\star}$ indices, quantifying the degree of morphological disturbance in the resolved mass maps and used to select morphological mergers. The two panels present the results for our two alternative approaches to estimating asymmetry (Section~\ref{sec:MorphoClass} and Appendix~\ref{app:AsyCentroid}).
On the left, the asymmetry is calculated using the peak of the mass distribution as centre of rotation. On the right, the asymmetry is calculated with respect to the mass centroid. The coloured, background area provides the merger probability in different regions of these planes based on the MIRAGE simulations of interacting and isolated galaxies in \citet[][see Paper I for further details]{Perret+14}.  
Symbols with error bars are galaxies from our sample (FIR+mm detected objects only, for illustration). Large, coloured symbols highlight galaxies with merger probability $>$85\% and therefore classified as mergers in Section~\ref{sec:MorphoClass} (as in Paper I, a minimum asymmetry of $A>0.1$ is imposed for mergers): Orange -- galaxies with $>$85\% merger probability according to both panels; green (magenta) -- galaxies satisfying the $>$85\% merger criterion only when using $A_{\star, \rm  peak}$ ($A_{\star, \rm centroid}$), respectively. White symbols are galaxies classified as non-interacting.}
\end{figure*}
%=============================%

Galaxy stellar masses for the FIR+mm sources are those published in L18 and correspond to the SED-inferred stellar mass of the closest optical counterpart in the 3D-$HST$ catalogues \citep{Skelton+14} or in the $K-$band selected sample of \citet{Pannella+15}. 
3D-$HST$ stellar masses are used for the 3D-$HST$ MS sample.

For all galaxies in our combined sample we also derived stellar mass maps following the approach detailed in Paper I.
Briefly, we performed pixel-by-pixel SED fitting to the available $HST$ photometry, after homogenisation of the SNR and convolution to match the $H$-band resolution. 
For the SNR enhancement we applied an adaptive smoothing using the code \textsc{Adaptsmooth}\footnote{{\scriptsize http://www.arcetri.astro.it/$\sim$zibetti/Software/ADAPTSMOOTH.html}} \citep{Zibetti_et_al_2009}, which replaces original pixel fluxes with averages over circular areas that reach the desired SNR threshold. As in paper I, we ran \textsc{Adaptsmooth} on a stack of all ACS and WFC3 images to set the required degree of smoothing.
 %=============================%
%==                FIGURE 4                    ==%
%=============================%

\begin{figure*}
\begin{center}
\includegraphics[width=0.93\textwidth]{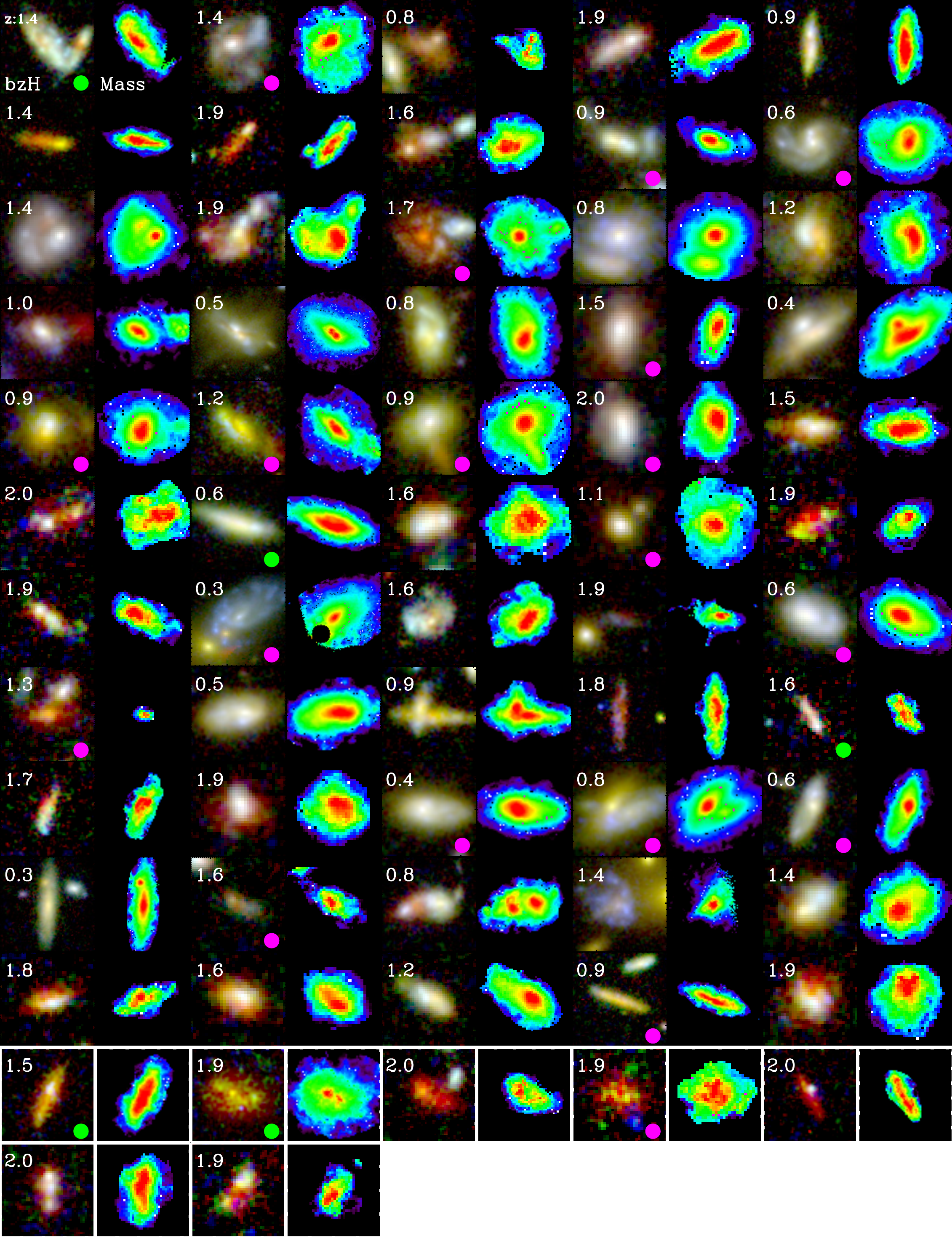}
\end{center}
\caption{\label{fig:MergerExamples} $B_{435}$- $z_{850}$-$H_{160}$ composite images and mass maps for galaxies classified as mergers according to the morphological classification scheme in section \ref{sec:MorphoClass}. Galaxies marked by a green (magenta) dot have a $>$85\% merger probability only when using $A_{\star, \rm  peak}$ ($A_{\star, \rm  centroid}$), respectively.
Stamp images with a white frame (bottom rows) highlight galaxies with low SNR, for which the merger classification is less certain. Galaxy redshifts are indicated in the upper left corner of each $B_{435}$- $z_{850}$-$H_{160}$ image. Stamp sizes are 2$\times$ the galaxy Kron radius.
}
\end{figure*}
%=============================%

We then fitted the adaptively smoothed pixel SEDs with \textsc{LePhare} \citep{Arnouts+99,Ilbert+06} using a set of \citet{Bruzual_Charlot_2003} templates with delayed, exponentially declining star formation histories (delayed $\tau$-models).
The characteristic time scale $\tau$ varied between 0.01 and 10\,Gyr in
22 logarithmically spaced steps and template ages were chosen between 100\,Myr and the age of the Universe at the given redshift. 
We corrected for internal dust extinction assuming a \citet{Calzetti+00} law with $E(B-V)$ ranging between 0 and 0.9 mag. The maximum $E(B-V)$ value corresponds to a dust attenuation in the far-UV of $A_{FUV}{\simeq}$9\,mag, which is suitable also for the SB regime (e.g., \citealt{Meurer+99,Overzier+11,Nordon+13}, with the exclusion of the most extremely obscured SBs/SB-regions which can reach  $A_{FUV}{>>}10$,  \citealt{Lester+90,Scoville+97,Engelbracht+98,Genzel+98,Puglisi+17}).
When generating the maps, any neighbouring galaxy that is at a different redshift than the primary galaxy in our sample -- i.e., is not a spectroscopic/photometric companion as described below -- was masked using the \textsc{SExtractor} segmentation map.
Total galaxy masses derived by summing up individual pixels in the resolved mass maps agree very well with the integrated stellar masses, with an rms error of $\sim$0.15\,dex (see also Paper I, Appendix A.1).

%%%%%%%%%%%%%%%%%%%%%%%%%%%%%%%%
%%%%%%%%     SEC.  CLASSIFICATION    %%%%%%%%%
%%%%%%%%%%%%%%%%%%%%%%%%%%%%%%%%
\section{Merger Classification}\label{sec:mergerClass}

\subsection{Morphologically selected mergers}

\subsubsection{Morphological classification on the mass maps}\label{sec:MorphoClass}
To morphologically classify our galaxies into mergers and non-interacting systems we follow the methodology of Paper I. To summarise, the merger classification is based on the degree of asymmetry/clumpiness of the galaxy as quantified by a combination of the Asymmetry index ($A$) \citep{Conselice_2003,Zamojski+07} and the normalized second-order moment of the $20\%$ brightest (most massive in our case) pixels,  $M_{20}$  \citep{Lotz+04}

The novelty of our approach lies in the fact that these structural parameters are measured on resolved mass maps, not broad-band images. 
We henceforth denote asymmetry and $M_{20}$ indices measured on mass maps as $A_{\star}$ and $M_{20, \star}$. As shown in Paper~I and briefly summarised in Section~\ref{app:HbadComp}, this method reduces the contamination from clumpy, non-interacting galaxies in the merger sample even with respect to a classification performed on NIR imaging ($H$-band). This is particularly important at $z{\gtrsim}1$ where the population of clumpy disc galaxies becomes more numerous \citep[e.g.][]{Cowie+95,Elmegreen+07,Forster_Schreiber+09,Guo+12,Guo+15,Murata+14}. 

In Paper I, we used the Merging and Isolated high redshift Adaptive mesh Galaxies (MIRAGE) hydrodynamic simulation from \citet{Perret+14} to generate artificial mass maps at the depth of our observations and estimated the ratio of merging vs. normal galaxies populating different regions of the $A_{\star}$--$M_{20, \star}$ plane.
We use this information to associate a merger probability to each galaxy in our sample, as illustrated in Fig.~\ref{fig:Class}. As a modification to the method presented in Paper I, we here use the asymmetry index calculated with respect to both the peak ($A_{\star, \rm  peak}$) and the centroid of the mass distribution ($A_{\star, \rm  centroid}$), and we classify as a merger any galaxy lying in a region with merger probability $>$85\% using either definition. 
This threshold was chosen as the optimal balance between completeness and purity of our classification.
The reason for including the alternative asymmetry estimate is that it is more sensitive to low mass ratio mergers (1:6), as outlined in Appendix \ref{app:AsyCentroid}, allowing us to increase the completeness of our sample by $\sim$15\% at these mass ratios. 
In comparison to $A_{\star, \rm  peak}$, the $A_{\star, \rm  centroid}$ index is generally more sensitive to a lopsided mass distribution, than to galaxy clumpiness (see, e.g., Fig.~\ref{fig:MergerExamples}). Therefore, using this extra parameter results in the inclusion of mergers that would otherwise be missed.

We show in Fig.~\ref{fig:MergerExamples} images of all galaxies classified as mergers in the FIR+mm and 3D-$HST$ MS samples.
We note that, like standard single-band morphological classifications, our mass-based classification scheme becomes less certain for galaxies close to the surface brightness limit of the observations where the mass maps become noisy (we estimate that reliable mass maps can be obtained for galaxies with surface brightnesses down to $\mu(H) \sim$26.5 mag\, arcsec$^{-2}$, see fig. 3 in Paper I). 
Although we performed an adaptive smoothing of the images prior to pixel-by-pixel SED fitting, the SNR can remain low even after smoothing for the faintest galaxies in our sample.
We thus flag galaxies morphologically classified as mergers but with average pixel SNR$>$3  in less than four optical--NIR bands (highlighted with white boxes in Fig.~\ref{fig:MergerExamples}, roughly 10\% of all morphological mergers). 
When computing merger fractions we will conservatively exclude these uncertain cases from the merger sample, but show with error bars how these fractions would change when including them. 

\subsubsection{Comparison with H-band classification} \label{app:HbadComp}	

Comparing the performance of the mass-based merger classification with respect to a standard $H$-band classification was the aim of Paper I, where we extensively tested our approach at different depths. While we do not wish to repeat that analysis here and refer the interested reader to our previous paper for an in-depth discussion, we review the key aspects to enable comparison with other studies.

\begin{figure}
\begin{center}
\includegraphics[width=0.48\textwidth]{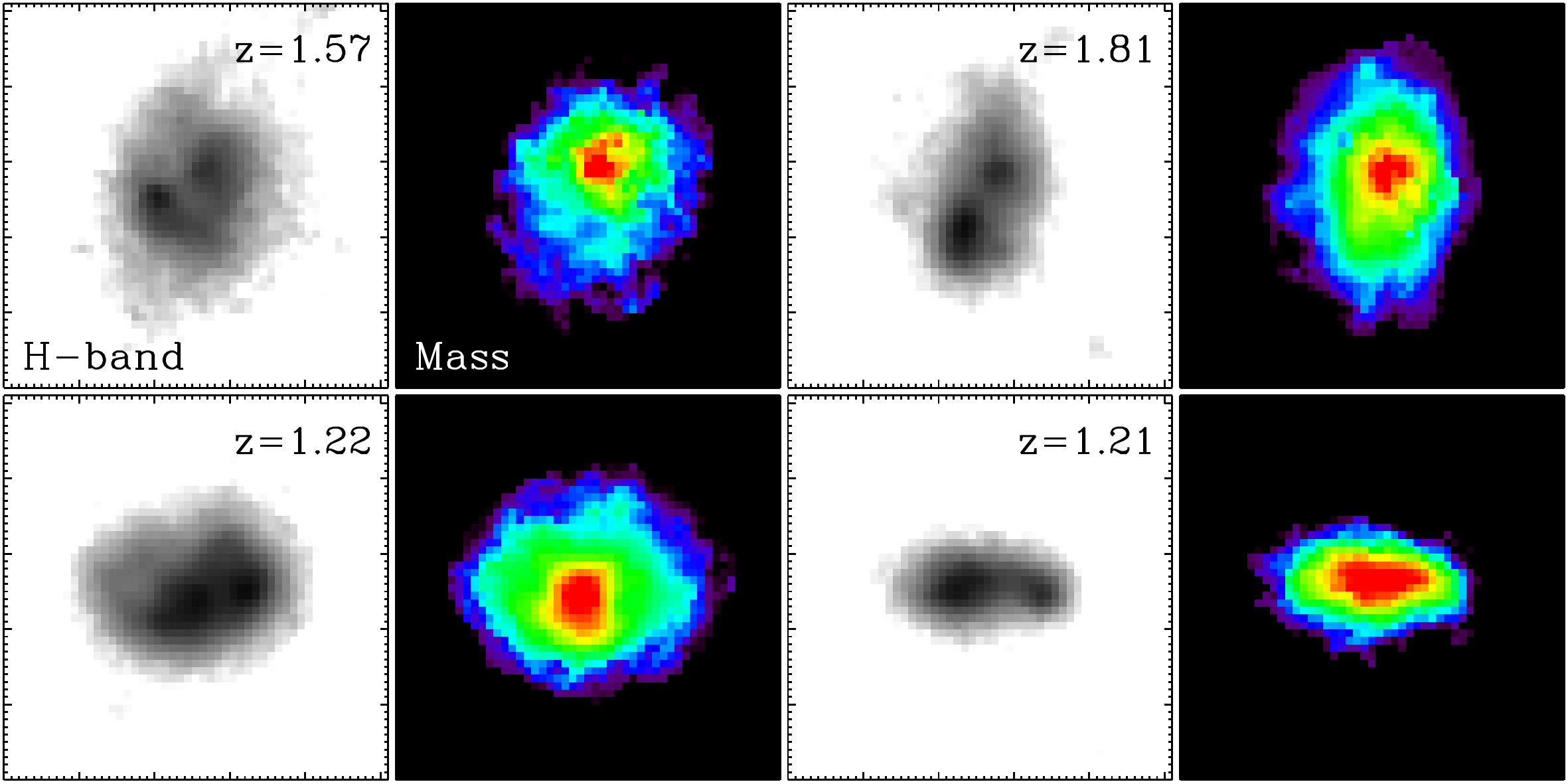}
\end{center}
\caption{\label{fig:FalseMergers} Examples of galaxies with a clumpy/disturbed H-band appearance but a smooth mass map. Images are 3$\arcsec$ a side.}
\end{figure}

If we apply our classification scheme to the $H$-band images, rather than the mass maps, we find that about 30\% of these $H$-band classified mergers display a smooth, disc-like mass distribution and, consequently, are \emph{not} classified as mergers using the mass maps. A few examples of such cases are shown in figure \ref{fig:FalseMergers}.
This is in line with our findings in Paper I that $H$-band-based classifications can have $\sim$40\% contamination from clumpy, non-interacting galaxies at the typical GOODS+CANDELS depth and is consistent with previous finding that mass maps are intrinsically smoother than the H-band images \citep{Wuyts+12}.
%=============================%
%==                FIGURE 5                    ==%
%=============================%
%=============================%
\subsection{Close spectro-photometric pairs}\label{sec:spec-photo_pairs}

The classification based on the asymmetry of the mass map will fail to identify mergers/interactions in which the two galaxies are still far apart and hence appear as separate, unperturbed systems.
In order to include earlier merger stages in our analyses, we also searched for pair systems among galaxies not classified as morphological mergers. 
We note that strong morphological disturbances can develop also right after the first pericentric passage, when the two galaxies are still separated. Galaxies in this configuration  satisfying our morphological criterion will be included in the sample of morphologically selected mergers, which therefore might be composed of some early phase interactions.  For simplicity, we will however refer to morphologically selected mergers as ``late-stage" mergers and pair systems as ``early-stage" mergers.
This section details the techniques we employ to find galaxy pairs.

In calculating the pair statistics, we rely on the 3D-HST spectroscopic and photometric catalogues \citep{Momcheva+16,Skelton+14}, which have  a high completeness.
For convenience, we refer to any galaxy in our final combined sample as a ``primary" galaxy, without any consideration to the mass difference with respect to its neighbouring, or ``secondary", galaxy. 
We also indicate the (spectroscopic or photometric) redshift and related uncertainty of the primary galaxy with $z_{p}$ and $\sigma_p$, and  those of the secondary with $z_{s}$ and $\sigma_s$. 

We start by searching in the 3D-HST catalogue for any secondary galaxy with an angular separation in arcseconds that would correspond to a physical projected separation $r_p$\,$\le$\,30kpc, at the redshift $z_{p}$ of the primary galaxy. 
If such a neighbour is found we then distinguish two cases.

Case (1): a spec-$z$ is available for both primary and secondary galaxy. In this case we classify the system as close spectroscopic pair if $\Delta$\,v=$c|z_{p}-z_{s}|/(1+z_{p})$\,$\le$500\,km/s.
This criterion is widely used in the literature and it has been shown that the majority of pairs with these spatial separations and relative velocities will eventually merge in about 1\,Gyr \citep[e.g.][]{Patton+2000,Kitzbichler_White_08}. 

Case (2): at least one of the two galaxies has no spectroscopic information, but only a photometric redshift. For these pairs we need to adopt a more probabilistic approach that takes into account the distribution in redshift space of the two galaxies and the redshift dependence for the  probability of being in a pair system. 
Here we followed the method outlined in \citet{LopezSanjuan+10}, to which we also refer for further details. We assign to each galaxy a Gaussian redshift probability distribution; e.g., for the primary galaxy this is: 

\begin{equation}
P_{\rm p}(z)= \frac{1}{\sqrt{2\upi}\sigma_{\rm p}} \exp{\left[-\frac{(z-z_{\rm p})^2}{2\sigma_{\rm p}^2}\right]} \, , 
\end{equation}
where $z_{\rm p}$ is the primary galaxy spec- or photo-$z$ and the standard deviation is given by either the typical spec-$z$ accuracy, $\sigma_{\rm p}=\sigma_{\rm spec}$\,$\simeq$\,0.003$(1+z)$ \citep{Brammer+12} or the photometric redshift uncertainty, $\sigma_{\rm p}=\sigma_{\rm phot}$, as provided in the 3D-HST catalogue.
An equivalent distribution,  $P_{\rm s}(z)$, holds for the secondary galaxy.
 
The probability that the two galaxies are in a pair system is then given by: 
\begin{equation}\label{eq:MergProb_photoPairs}
\zeta_{\rm pair}(z)=\alpha P_{\rm p}(z)\int^{z_{\Delta \rm v_{max} }}_{z_{\Delta \rm v_{min} }}P_{\rm s}(z')dz' \, ,
\end{equation}
where $\alpha$ is a normalisation constant and $z_{\Delta \rm v_{max} }$ and $z_{\Delta \rm v_{min} }$ are defined by imposing that the secondary galaxy satisfies the condition for being a close companion at any given redshift $z$, $\Delta$\,v=$c|z-z_{\Delta \rm v_{min/max}}|/(1+z)$\,$\le$500\,km/s. 

We assume that the two galaxies can be a close pair, and therefore have $\zeta_{\rm pair}(z)$\,$\neq$\,0, in the redshift range $[z_{\rm l},z_{\rm u}]$, where the boundaries $z_{\rm l}$ and $z_{\rm u}$ are given either by the 
intersection region $[z_{\rm p}-3\sigma_{p}, z_{\rm p}+3\sigma_{p}]$\,$\cap$\,$[z_{\rm s}-3\sigma_{s}, z_{\rm s}+3\sigma_{s}]$ or the redshift limits over which the two galaxies would still lie at a projected distance $r_p$\,$\le$30kpc, whichever is most stringent.
We define the normalisation constant $\alpha$ such that: 
\begin{equation}
\int^{z_{\rm u}}_{z_{\rm l}}\zeta_{\rm pair}(z)dz=\int^{z_{\rm u}}_{z_{\rm l}} P_{\rm p}(z) dz \, \times \int^{z_{\rm u}}_{z_{\rm l}} P_{\rm s}(z) dz \,.
\end{equation}
With this definition of $\alpha$, the integral of $\zeta_{\rm pair}(z)$ represents the fractional value of the primary galaxy that is in a pair system with the secondary.
To better clarify the procedure, Appendix \ref{app:PhotometricPairs} presents some examples of the derivation of $\zeta_{\rm pair}(z)$ for different combinations of spectro-photometric pairs.

Pushing a step further this formalism, we can associate a $\zeta_{\rm pair}(z)$ function also to mergers classified using the morphological method in Section~\ref{sec:MorphoClass} and to the spectroscopic pairs described in case (1). By definition, these galaxies are merging at any redshift in which they exist and for this reason have a merger probability distribution equal to their redshift distribution, i.e., $\zeta_{\rm pair}(z)$\,=\,$P_{\rm p}(z)$.

A number of works in the literature have used a simpler photometric redshift difference threshold to identify close pairs, rather than following the probabilistic approach described above \citep[e.g.,][]{Kartaltepe+07,Ryan+08,Bundy+09,Man+16}.
We have tested how our results would vary if we identified as a pair any galaxy with a neighbour having a redshift difference $\Delta z$\,$\leq$\,$0.1\times (1+z_{p})$.
Adopting this alternative definition would lead to an overall increase of about a 5\% of the merger fraction, but we found no significant changes in our findings. 

In the following we assume that all close pairs (after taking into account the statistical correction described below) will eventually merge. Some of these pairs, however, could be weakly bound or unbound.
The impact on our results of pairs that do not merge is briefly discussed in Section~\ref{sec:MergerTimes} in the context of merger time-scales. The other results are instead strongly driven by the morphologically selected mergers, such that the presence of these pairs has minimal effect on our findings.

%%%%%%%%%%%%%%%%%%%%%%%%%%%%%%%%
%%%%%%%%       SEC.  CORRECTIONS    %%%%%%%%%
%%%%%%%%%%%%%%%%%%%%%%%%%%%%%%%%

\section{Statistical corrections to the merger fraction}\label{sec:Corrections}

\subsection{Minimum mass ratio} 
For interpreting merger fraction measurements, it is important to know down to which mass ratio our merger selection is sensitive.
The morphological selection technique developed in Paper I and described in Section \ref{sec:mergerClass} has been calibrated on morphologically disturbed mergers with mass ratios in the range ratio 1:1 to 1:6.3 and is sensitive to the immediate pre-coalescence phase ($\sim$300\,Myr before coalescence in the simulations).
In Appendix~\ref{app:AsyCentroid} we use our galaxy sample to generate a set of artificial mergers and quantify how the morphological selection completeness depends on the mass ratio.
Our selection is $\geq$70\% complete down to ratios of 1:4, $\sim$60\% down to ratios 1:6 and then decreases significantly below this value. We hence assume a mass ratio threshold of 1:6  for our selection from now on.
To consistently combine the morphologically identified mergers with those from the spectro-photometric selection, we restrict the close spectro-photometric pair sample to pairs with a ratio of 1:6.

This raises the question of whether, for this mass ratio, the catalogues used for pair selection are complete to companions. Given the mass limits described in Section~\ref{sec:SampleCompleteness}, a 1:6 mass ratio translates to a minimum companion galaxy mass of $\log(M_{\star}/\Msol)$\,=\,8.6,  9.1 and 9.4 for our three redshift bins.
We estimate that, with a depth of $H_{F160W}$=26.9 mag \citep{Grogin+11}, the 3D-$HST$/CANDELS catalogue is 90\% complete for masses $\log(M_{\star}/\Msol)$>8.3 up to redshift $z{=}0.6$,  for $\log(M_{\star}/\Msol)$>8.7 up to $z{=}1.2$ and $\log(M_{\star}/\Msol)$>9.0  up to $z{=}2.0$.
Incompleteness in the spectro-photometric pairs is therefore not an issue. With the method described in Section~\ref{sec:spec-photo_pairs}, in the redshift range 0.2${\leq}z{\leq}$2  we find a total of 16 pure spectroscopic pair systems with a mass ratio 1:6 and 22 spectro-photometric pairs with a mass ratio down to 1:6 and a non-negligible integral of $\zeta_{{\rm pair}}(z)$ ($\int \zeta_{{\rm pair}}(z)dz>$0.3).

\subsection{Corrections for spectro-photometric pairs} 

Beside companion completeness, there are two other factors that can affect the spectro-photometric pair estimate and therefore need to be taken taken into account. 

First, while the probabilistic approach described in Section~\ref{sec:spec-photo_pairs} accounts for photo-$z$ uncertainties, these can be quite large and physically unassociated, projected pairs could still bias the calculation of the merger fraction.    
To estimate the contamination by these chance projections we followed the approach of \citet{Man+16} and generated 11 realizations of our galaxy catalogs in which we randomized the R.A. and Dec. position of all galaxies, while leaving all other properties unchanged. We then calculated the average pair number, $\bigg\langle \sum \int^{z_2}_{z_1}\zeta_{\rm pair}(z, {\rm random}) \bigg\rangle$, in these random samples and subtracted it from the original measurement.

Second, as shown in panel \emph{(e)} of Fig.~\ref{fig:Selection}, the fraction of galaxies with spec-$z$s decreases substantially at $z{\gtrsim}1.2$, due to the difficulty of obtaining reliable spectra at these redshifts. In our highest redshift bin $\sim$40\% of the galaxies in the combined FIR+mm and MS sample have a spec-$z$ measurement, compared to $>$80\% in the first two bins.
A similar trend is observed in the distribution of spectroscopic redshifts in the 3D-$HST$ catalogue used to identify pair candidates (albeit starting from a lower overall spectroscopic completeness).
Consequently, the number of spectro-photometric pairs where at least one galaxy has a spec-$z$ decreases with redshift and in the highest redshift bin the pair identification heavily relies on photo-$z$'s, bearing the risk of introducing systematic effects when relating merger probabilities between redshift bins.

To asses the extent of this effect, we recalculated pair probabilities and fractions after replacing spec-$z$'s with photo-$z$s for a random subset of galaxies (both primary and secondary) in the first two redshift bins, until reproducing the ${\sim}40\%$ spectroscopic completeness of the $1.2{\leq}z{\leq}2$ bin. Merger fractions calculated in this way differ from the original estimate by only $\sim$10\% (consistent findings are presented in \citet{LopezSanjuan+10} for similar spectroscopic fractions). Even if the spectroscopic coverage is only partial, it is therefore not necessary to apply corrections for this effect. 

\subsection{Correction matrix for morphological mergers} \label{sec:ComplMatrix}

We can also derive statistical correction factors for the mergers classified via the morphological technique.
The \emph{true} number of morphological mergers in our sample is mapped into an observed number of \emph{classified} mergers which is not necessarily identical to the intrinsic value. The \emph{completeness} of our selection, i.e., the fraction of true mergers identified as such, and the \emph{contamination}, i.e., the fraction of morphologically selected mergers, that in truth are not interacting and are therefore misclassified, both contribute to differences between these numbers.

We can use tests on simulated and real galaxies to derive a correction scheme enabling us to recover the intrinsic morphological merger fraction. 
For simplicity, in the following we label and refer to non-interacting galaxies as ``discs". This is a purely notational choice and has no physical implication for the actual disc prominence in the non-interacting sample, which will also contain bulge-dominated galaxies.  

We start by defining a ``completeness/contamination" matrix $\mathbf{C}$ as follows:
\begin{eqnarray} \label{eq:MatrixC}
\notag
\begin{bmatrix}
  N_{\rm M} \\
     N_{\rm D}     
\end{bmatrix}
  &= &  
 \mathbf{C} \times
\begin{bmatrix}
  N_{\rm True,M} \\ 
  N_{\rm True,D}      
\end{bmatrix} =
  \\
  \notag
  &=&
   \begin{bmatrix}
 C_{11} & C_{12} \\
  C_{21} & C_{22}  
  \end{bmatrix} 
  \times
\begin{bmatrix}
  N_{\rm True,M} \\ 
  N_{\rm True,D}      
\end{bmatrix} 
  = 
  \\
  & = &
\begin{bmatrix}
  P_{\rm M\rightarrow M}    &   P_{\rm D\rightarrow M}   \\
   P_{\rm M\rightarrow D}    &   P_{\rm D\rightarrow D}   
\end{bmatrix}
\times 
\begin{bmatrix}
  N_{\rm True,M} \\
   N_{\rm True,D}       
\end{bmatrix} \,,
\end{eqnarray}
where $N_{\rm M} $ and $N_{\rm D}$ are the number of galaxies classified as mergers and discs in our sample. $N_{\rm True,M}$ and $N_{\rm True,D}$ are the real number of mergers and discs. The terms inside the matrix $C$ describe the probability of a true merger or disc being classified as either of the two classes. For example, $C_{21}=P_{\rm M\rightarrow D}$ is the probability for a true merger to be classified as disc instead.
We can estimate $N_{\rm True,M}$ -- and therefore  recover the intrinsic merger fraction -- by inverting $\mathbf{C}$:
\begin{eqnarray*}
\begin{bmatrix}
 N_{\rm True,M} \\
 N_{\rm True,D}       
\end{bmatrix}
 &= & \mathbf{C^{-1}} \times 
 \begin{bmatrix}
  N_{\rm M} \\  
  N_{\rm D}     
\end{bmatrix} 
= \\
&=& 
\frac{1}{C_{11}C_{22}-C_{12}C_{21}}
 \begin{bmatrix}
  C_{22} & -C_{12} \\
  -C_{21} & C_{11}  
\end{bmatrix}
\times
 \begin{bmatrix}
  N_{\rm M} \\  
  N_{\rm D}     
\end{bmatrix}\, ,
\end{eqnarray*}
where the transformation from ($N_{\rm M}$,$N_{\rm D}$) to ($N_{\rm True,M}$,$N_{\rm True,D}$)   via $\mathbf{C^{-1}}$  is number conserving. 

This leads to:
\begin{equation}\label{eq:Ntrue}
N_{\rm True,M} = \frac{C_{22} N_{\rm M} - C_{12} N_{\rm D}}{C_{11}C_{22}-C_{12}C_{21}} \,.
\end{equation}
The full formalism for the derivation of $\mathbf{C}$ is not complex, but somewhat lengthy. Readers not interested in these details can proceed directly to Equation~\ref{eq:CorrectionValues} and the subsequent text, where we provide an interpretation of Equation~\ref{eq:Ntrue}.

Two main factors define the completeness and contamination in our selection, hence the probability terms in matrix $\mathbf{C}$. 
First, the classification scheme inherently carries a certain degree of incompleteness as well as mixing between the disc and merger populations  which is determined by the chosen classification threshold (i.e., 85\% merger probability in our case). This reflects the ``intrinsic" overlap of the two classes of galaxies and even an ideal, noise-free sample would be affected by it. 
Second, in real data S/N effects introduce an additional source of confusion between mergers and discs, scattering the two classes across the  $A_{\star}$-$M_{20, \star}$ plane. 

We treat these two factors as independent terms, such that they can be parametrized by two separate matrices  $\mathbf{P_{C}}$ and $\mathbf{P_{N}}$ representing the effects of classification and of noise. These matrices map the true number of mergers and discs into classified discs and mergers as follows: 
\begin{eqnarray}\label{eq:PnPc}
\begin{bmatrix}
  N_{\rm M} \\  
  N_{\rm D}     
\end{bmatrix}
  & = &
 \mathbf{P_{N}} \times  \left( \mathbf{P_{C}}  \times
\begin{bmatrix}
  N_{\rm True,M} \\
   N_{\rm True,D}        
\end{bmatrix} 
\right) =
  \\
  & = &
  \notag
  \begin{bmatrix}
   P_{\rm M, class \rightarrow M, noise}    &   P_{\rm D, class \rightarrow M, noise}   \\
   P_{\rm M, class \rightarrow D, noise}    &   P_{\rm D, class \rightarrow D, noise}   
\end{bmatrix} 
 \times \\
& \times &
\left(
\notag
\begin{bmatrix}
   P_{\rm M, true \rightarrow M, class}    &   P_{\rm D, true\rightarrow M, class}   \\
   P_{\rm M, true \rightarrow D, class}    &   P_{\rm D, true\rightarrow D, class}   
\end{bmatrix}
\times 
\begin{bmatrix}
  N_{\rm True,M} \\
   N_{\rm True,D}        
\end{bmatrix}
\right) \,. 
\end{eqnarray}
In the equations above $P_{\rm M, true \rightarrow M, class}$ is the probability that a \emph{true} merger is classified as a merger purely on account of classification threshold choices and $P_{\rm M, class \rightarrow M, noise}$ is the probability that a galaxy classified as a merger in noise-free circumstances retains its classification as a merger after accounting for noise in the data. 
Similar expressions hold for the crossover terms and for the disc population. 

The matrix $\mathbf{P_{C}}$ quantifying the degree of mixing inherent to the classification can be inferred from the artificial simulations we used to calibrate the classification scheme itself\footnote{The simulations were processed to reproduce the S/N in the actual observations and therefore the two factors discussed above are somewhat intertwined. However, more complex S/N effects present in the data (e.g., wavelength-dependent impact of noise on the SED fitting),  could not be fully reproduced in the simulations and hence we use the comparison of the real data at different depth as a way of quantifying this additional term.}. The probability terms in $\mathbf{P_{C}}$ are simply given by the fraction of all simulated mergers and discs that were classified or misclassified with our technique.
For noise-related effects and the matrix $\mathbf{P_{n}}$, we rely instead on tests performed on the real data. In Paper I, we have derived mass maps for a set of galaxies in the Hubble Ultra Deep Field (HUDF) at two different depths: using the very deep HUDF and HUDF12 images or instead the shallower CANDELS and GOODS imaging, which has the same depth as our GOODS-N sample.
By applying the morphological classification scheme to both sets of maps, we can estimate how S/N-effects scatter galaxies and blur the classification in real data. The fraction of galaxies with consistent or discrepant classes at the GOODS-depth with respect to the deeper HUDF classification correspond to the terms in matrix $\mathbf{P_{N}}$. 

Comparing Eq.~\ref{eq:PnPc} with Eq.~\ref{eq:MatrixC} we see that: 
\begin{eqnarray*}
 \mathbf{C} &=&  \begin{bmatrix}
 C_{11} & C_{12} \\
  C_{21} & C_{22}  
  \end{bmatrix} = 
   \mathbf{P_{N}} \times   \mathbf{P_{C}}  = \\
 &= & 
 \begin{bmatrix}
  P_{N,11}P_{C,11}+ P_{N,12}P_{C,21} & P_{N,11}P_{C,12}+ P_{N,12}P_{C,22} \\     
  P_{N,21}P_{C,11}+ P_{N,22}P_{C,21} & P_{N,21}P_{C,12}+ P_{N,22}P_{C,22} \\      
\end{bmatrix} \,,
\end{eqnarray*}
where, as an example, $P_{N,11}=P_{\rm M, class \rightarrow M, noise}$.
 From the analysis described above, we derive the following values for the correction terms:
\begin{equation}\label{eq:CorrectionValues}
 \mathbf{C} =
 \begin{bmatrix}
 0.55 & 0.05 \\
 0.45 & 0.95
  \end{bmatrix}
  \qquad {\rm and } \qquad
  \mathbf{C^{-1}} =
 \begin{bmatrix}
 1.90 & -0.10 \\
 -0.90 & 1.10
  \end{bmatrix}\,.
\end{equation}

We can understand the nature of our correction as follows.
When the classified number of mergers ($N_M$) is significantly larger than the number of classified discs ($N_D$), Equation~\ref{eq:Ntrue} simplifies to $N_{\rm True,M}\xrightarrow{N_M{\gg}N_D} C_{22} N_{\rm M}/(C_{11}C_{22}-C_{12}C_{21}) \simeq N_{\rm M}/C_{11}$ and the dominant contribution to the correction arises from matrix element $C_{11}{=}P_{\rm M\rightarrow M}{\equiv}$55\%, which represents the completeness of our merger selection. This overall merger completeness results from the combination of classification-driven incompleteness ($\sim$30\%\footnote{The classification-driven incompleteness estimated here, from simulations, for the completeness/contamination matrix $\mathbf{C}$ agrees within 10\% with the incompleteness inferred using artificially generated mock merger images (see Appendix \ref{app:AsyCentroid}).}) and noise effects (adding a further $\sim$15\%). 
Conversely, although only a small fraction of discs suffers from misclassification ($C_{12}{=}P_{\rm D\rightarrow M}{\equiv}$5\%), their contribution is non-negligible once $N_D{\gtrsim}N_M$ and the correction factor $C_{12}N_D$ in Equation~\ref{eq:Ntrue} must be taken into account.
For strongly disc-dominated samples the measured merger count may in fact be dominated by misclassified discs, resulting in a merger fraction consistent with zero.
All morphological merger fractions reported from now on (Figs~\ref{fig:FracMerger}-\ref{fig:time-scales}) have been corrected to intrinsic numbers following Equation~\ref{eq:Ntrue}.

\subsection{Merger fraction}

Combining all the information derived in the preceding sections,
we can now write the fraction of merging galaxies over all galaxies $N_{\rm tot}$ in a redshift interval $[z_1, z_2]$ as: 
\begin{equation}\label{eq:f_merge}
 f_{\rm merg}|_{z_1} ^{z_2} = \frac{N_{\rm zPair}+N_{\rm zpPairs}+N_{\rm morph}}{N_{\rm tot}} \, .
\end{equation}
The three terms in the numerator of Equation~\ref{eq:f_merge} are the number of spectroscopic pairs ($N_{\rm zPair}$), of spectro-photometric pairs corrected for chance projection ($N_{\rm zpPairs}$) and of morphologically classified mergers corrected to the true number ($N_{\rm morph}$). 

Specifically, following our formalism, we can write the number of spectroscopic pairs -- which we recall have $\zeta_{\rm pair}(z)$\,=\,$P_{\rm p}(z)$ -- as:
\begin{equation}
N_{\rm zPair}=\sum\limits_{j, \rm zPair} \int^{z_2}_{z_1}\zeta_{j, \rm pair} (z) dz = \sum\limits_{j, \rm zPair} \int^{z_2}_{z_1} P_j(z)dz \, .
\end{equation}
The number of spectro-photometric pairs is given by:
\begin{equation}
N_{\rm zpPairs}=\sum\limits_{j, \rm zpPair} \int^{z_2}_{z_1}\zeta_{j, \rm pair} (z) dz - \bigg\langle \sum \int^{z_2}_{z_1}\zeta_{\rm pair}(z, {\rm random}) \bigg\rangle \, ,
\end{equation}
and the intrinsic number of morphologically disturbed galaxies is:
\begin{equation}
N_{\rm morph}=\frac{C_{22}  \sum\limits_{j, \rm Mergers} \int^{z_2}_{z_1} P_j(z)dz - C_{12} \sum\limits_{j, \rm Disks} \int^{z_2}_{z_1}  P_j(z)dz}{ C_{11}C_{22}-C_{12}C_{21}} \, .
\end{equation}
The total number of galaxies in the denominator of  Equation~\ref{eq:f_merge} is simply:
\begin{equation}
N_{\rm tot}=\sum\limits_i \int^{z_2}_{z_1} P_{i}(z)dz \, .
\end{equation}

We note that the majority of our galaxies have a $P(z)$ that is fully included in one of the redshift bins used in this study, resulting in a $\int^{z_2}_{z_1} P(z)dz$\,=\,1 in the numerator of Equation~\ref{eq:f_merge}. With this formalism, we can however correctly redistribute the contribution of the $\sim$20\% of galaxies that have a redshift distribution straddling more than one bin.

%=============================%
%==                FIGURE 6                    ==%
%=============================%
\begin{figure*}
\begin{center}
\begin{tabular}{cc}
\includegraphics[width=0.47\textwidth]{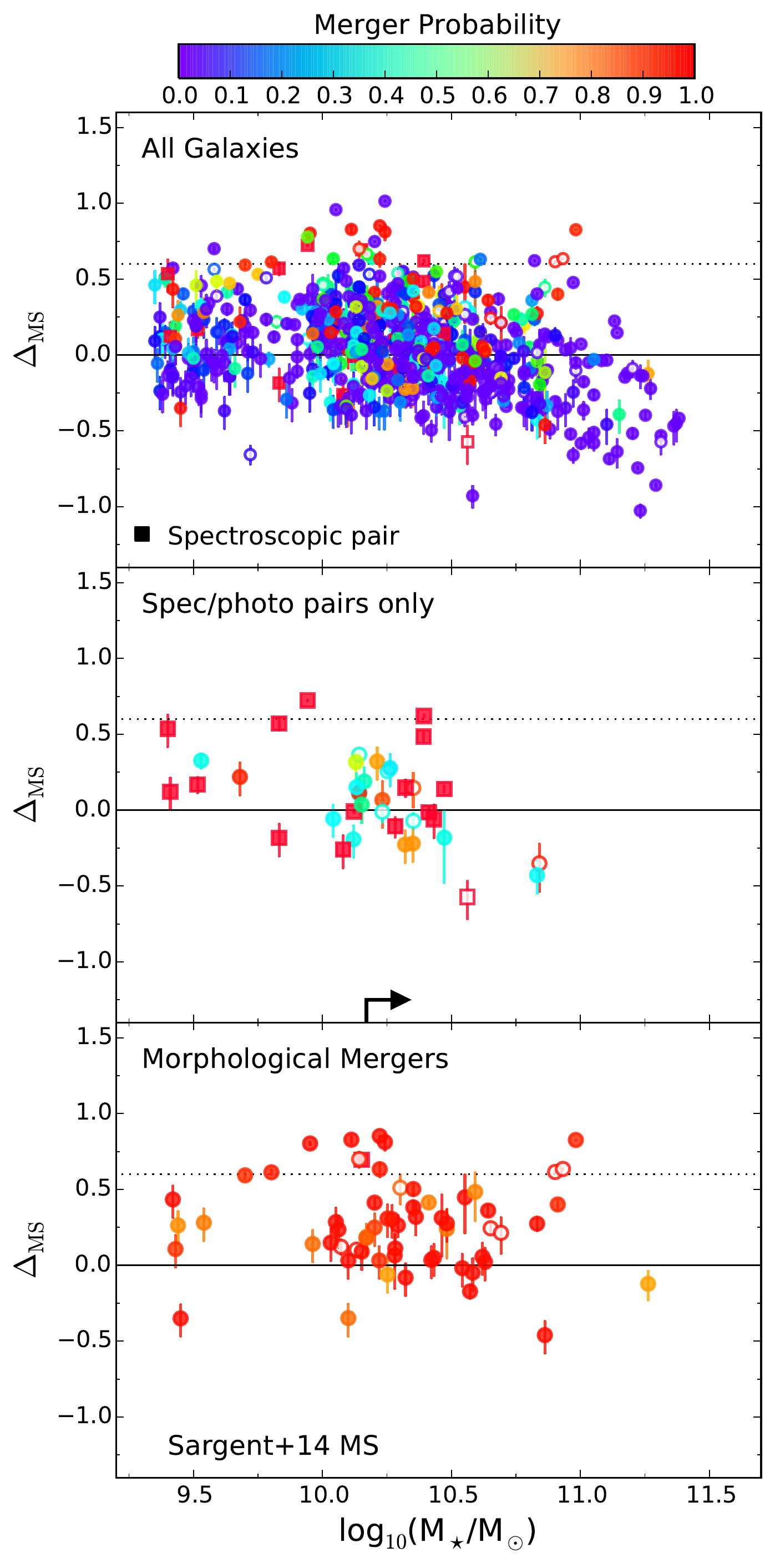} & 
\includegraphics[width=0.47\textwidth]{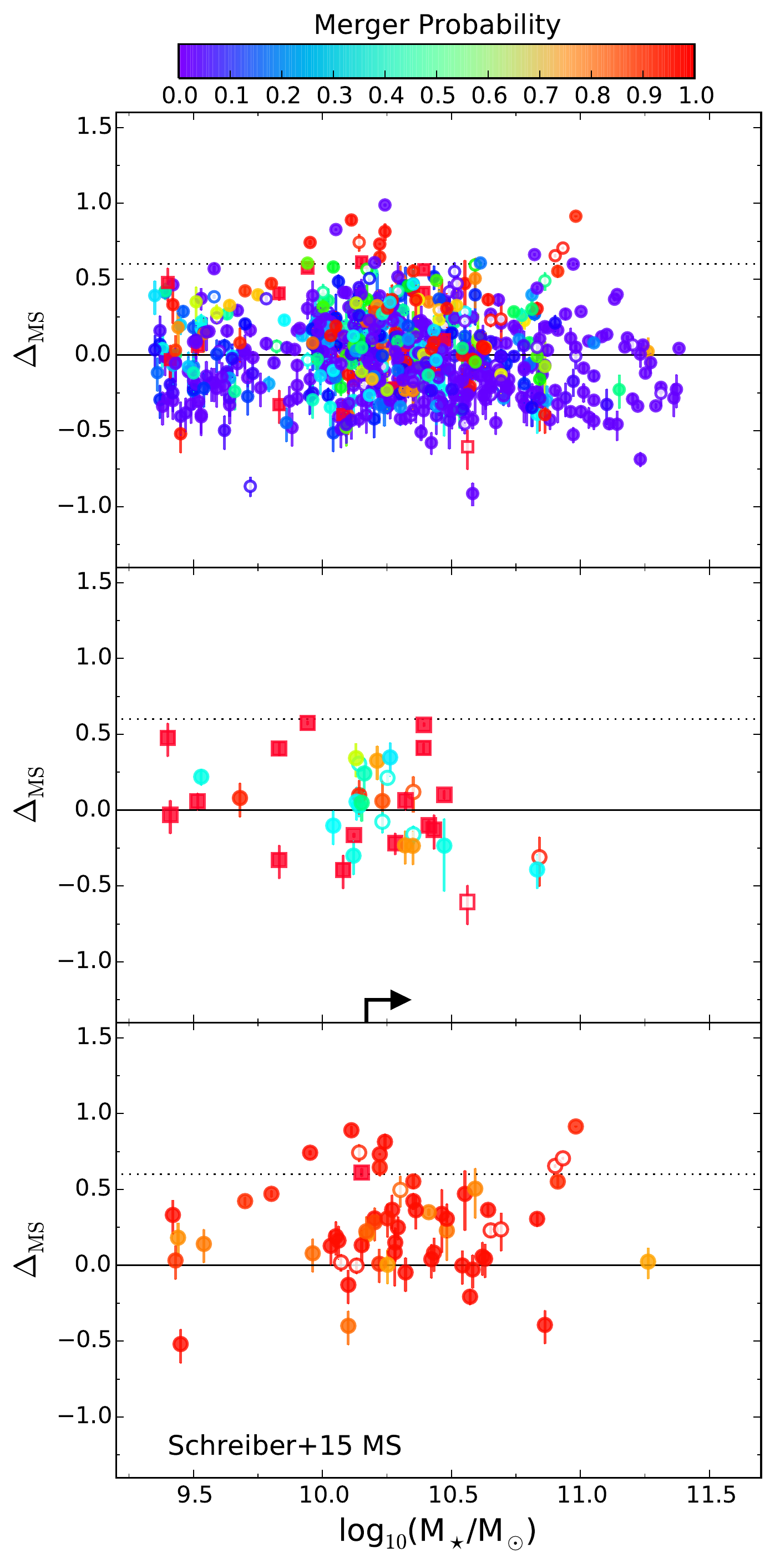}  \\
\end{tabular}
\end{center}
\caption{\label{fig:DeltaMS} \emph{Top:} distance $\Delta_{\rm MS}$ from the MS for galaxies in our sample, colour-coded by merger probability (see colour bar). The redshift-dependent locus of the MS is parametrized as in \citet[][left]{Sargent+14} or \citet[][right]{Schreiber+15}.
The merger probability of morphologically identified mergers is inferred from the position in $A_{\star}$-$M_{20, \star}$ space (max. merger probability between those calculated using either $A_{\star, \rm peak}$ or $A_{\star, \rm centroid}$). For galaxies in pairs it is given by the integral of the function $\zeta_{\rm pair} (z)$ in Section~\ref{sec:spec-photo_pairs} (spectroscopic pairs have a pair probability equal to 1 by definition).
Galaxies for which the IR/sub-mm photometry could suffer from blending (see Section~\ref{sec:PossibleBlends}) are shown with empty symbols. Dotted horizontal lines indicate a 4-fold SFR-enhancement with respect to the MS, i.e. the SB region.
\emph{Middle:} as above, but for spectro-photometric pairs. Only spectro-photometric pairs with non-negligible merger probability $\int \zeta_{\rm pair}(z) dz \ge 0.3$ are shown. Square symbols represent pure spectroscopic close pairs (where both primary and secondary galaxies have a spec-$z$). Black arrows show the stellar mass above which our sample is complete at all redshifts for MS galaxies.
\emph{Bottom:} distribution of morphologically identified mergers.
}
\end{figure*} 
%=============================%

%%%%%%%%%%%%%%%%%%%%%%%%%%%%%%%%
%%%%%%%%                  RESULTS              %%%%%%%%%
%%%%%%%%%%%%%%%%%%%%%%%%%%%%%%%%
\section{Results}\label{sec:Results}

\subsection{Merger fraction on the MS and among starburst galaxies}\label{sec:FracMerg}

We start our investigation of how major mergers affect star formation in galaxies by showing in Fig.~\ref{fig:DeltaMS} the distance from the MS for the entire sample of galaxies, this time highlighting with colours the merger probability for each galaxy, as either derived from its morphology in Section~\ref{sec:MorphoClass} or from the pair statistics in Section~\ref{sec:spec-photo_pairs}.
Results are presented for the two parametrizations of the MS \citep{Sargent+14,Schreiber+15}.
We also show separately the distribution of close pairs only (middle panel) and of galaxies that are in a late stage merger as probed by the morphological selection (bottom panel).
%=============================%
%==                FIGURE 7                    ==%
%=============================%
\begin{figure}
\begin{center}
\includegraphics[width=0.48\textwidth]{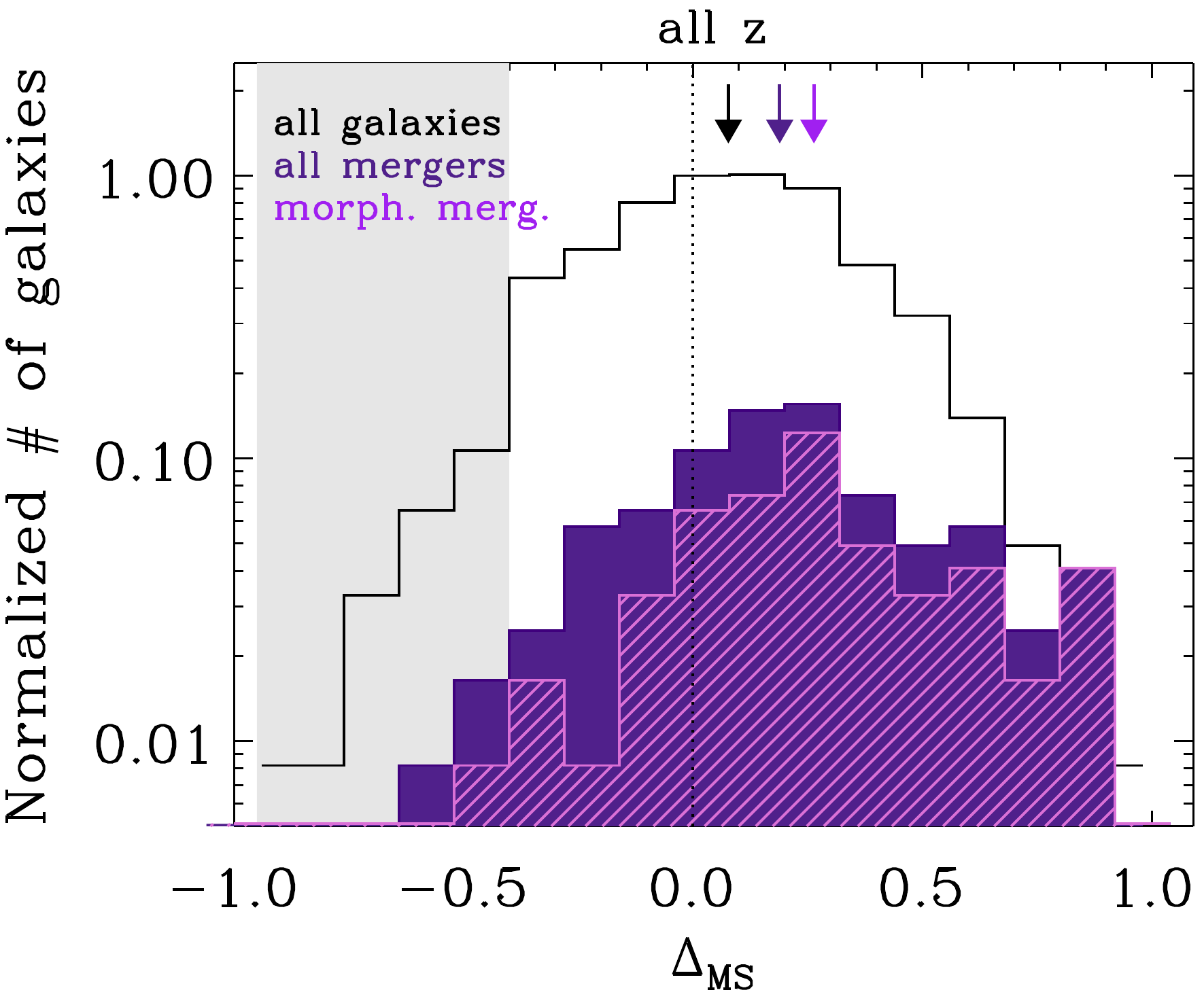}
\end{center}
\caption{\label{fig:HistDMS} Observed $\Delta_{\rm MS}$ distribution, adopting the \citet{Sargent+14} MS  parametrization, for our entire galaxy sample at $0.2{\leq}z{\leq}2.0$ (black histogram), for all major mergers shown in Fig~\ref{fig:DeltaMS} (morphological mergers+pairs, dark purple) and for morphological mergers only (light purple). The median $\Delta_{\rm MS}$ for the three galaxy samples are shown with colour-matched arrows. Curves are normalized to the total number of galaxies in the $\Delta_{\rm MS}$=0 bin (highlighted with dotted line). The shaded grey area indicates $\Delta_{\rm MS}$ values where our sample becomes incomplete. 
Although major mergers can be found at any $\Delta_{\rm MS}$, their distribution is skewed toward higher sSFR than the global population.}
\end{figure}
%=============================%

Where our sample selection is complete ($\Delta_{\rm MS}{\ge}-0.4$), merging galaxies are found over a wide range of $\Delta_{\rm MS}$. 
However, the relative contribution of merging galaxies to the overall population depends on the distance from the MS and the average merger probability increases when moving across the MS, as evident in the upper panels of Fig.~\ref{fig:DeltaMS}. This is more explicitly shown in Fig.~\ref{fig:HistDMS}, where we compare the distribution of  $\Delta_{\rm MS}$ for all galaxies at $0.2{\leq}z{\leq}2.0$ with that for mergers only.
Merging galaxies are shifted towards higher sSFR than the entire population; the median offset from the MS for all galaxies is $\big\langle \Delta_{\rm MS} \big \rangle{=}0.09_{-0.01}^{+0.01}$, whereas for all types of mergers combined we measure $\big\langle \Delta_{\rm MS} \big \rangle {=}0.21_{-0.04}^{+0.01}$. The difference is even larger when only considering morphologically selected mergers, namely $\big\langle \Delta_{\rm MS} \big \rangle{=}0.26_{-0.01}^{+0.04}$.

To further quantify this, we present the fraction of galaxies classified as mergers as a function of the distance from the MS in Fig.~\ref{fig:FracMerger}.
We show this for the entire sample and for the three separate redshift bins.
As discussed in Section~\ref{sec:MorphoClass}, uncertain morphological classification of low S/N galaxies could potentially affect the calculation of the merger fraction. 
In Fig.~\ref{fig:FracMerger} we show with white points the ``fiducial" merger fraction calculated by excluding these galaxies from the merger sample, but we use the coloured bar to indicate how the fraction would change if these galaxies were all genuine mergers.  In the size of these bars we also integrate the effect of removing galaxies for which the IR flux could be affected by blending (see Section~\ref{sec:PossibleBlends}) and the formal uncertainty on the merger fraction.
For completeness, we show in Appendix~\ref{app:FracnoCorrection} the same figure but   without applying any of the statistical corrections described in Section~\ref{sec:Corrections}.

The fraction of major mergers is roughly constant at values of $\sim$5--10\%  for galaxy that are within the scatter of the MS ($-0.3{\lesssim} \Delta_{MS} {\lesssim}0.3$ ), but starts to increase for  $\Delta_{MS}\gtrsim 0.3$. These trends are observed in all redshift bins and independently of the MS definition which is used.
Of those galaxies that are located in what is typically identified as the SB region, i.e., have  $\Delta_{MS}{\ge}0.6$, at least $\sim$70\% are found to be in a major merger.
The merger fraction increase is largely due to morphologically classified mergers, i.e., late-stage mergers, whereas the fraction of galaxy pairs stays roughly constant with only a modest increase at the highest $\Delta_{MS}$.

\subsection{Mass and redshift dependence of the MS merger fraction}
 \subsubsection{Mass dependence}
\begin{figure*}
\begin{center}
\includegraphics[width=0.453\textwidth,angle=90]{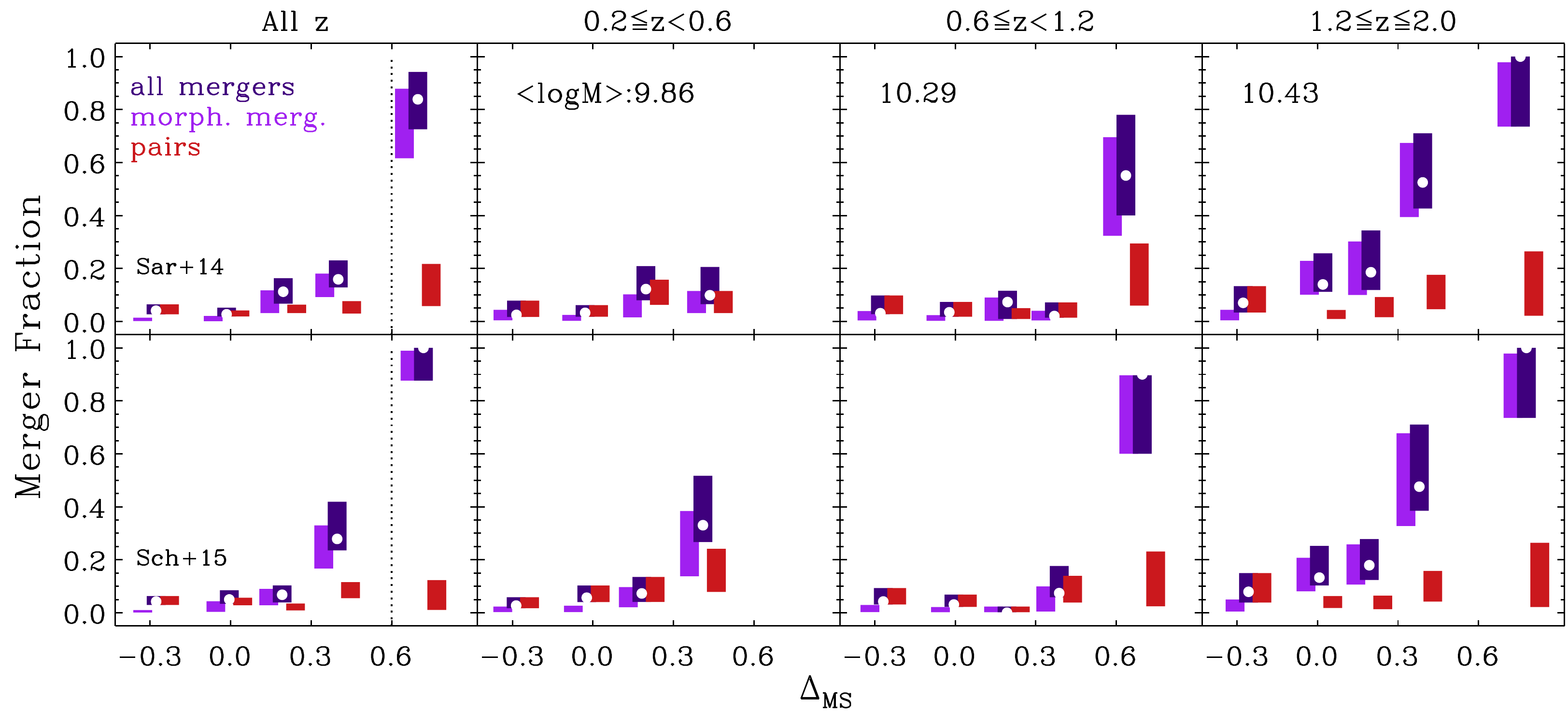}
\end{center}
\caption{\label{fig:FracMerger} Fraction of galaxies classified as mergers as a function of distance from the MS. Results are presented for the entire sample (``all z"), and separated by redshift bin.
For row 1 we adopt the MS parametrization of \citet{Sargent+14}, for row 2 that of \citet{Schreiber+15}.
We use boxes to represent the possible range of merger fractions measured when accounting for different sources of systematic uncertainties as discussed in the text (i.e., in/excluding galaxies with IR-blended photometry and galaxies with uncertain morphological classification) and for formal errors on the calculated fraction.
Dark purple bars show the total merger fraction, and light purple bars the fraction of morphologically selected mergers. Red bars correspond to the fraction of galaxies in pairs. 
 The white points mark the fiducial total merger fraction, calculated by excluding from the merger sample galaxies with an uncertain morphological selection. Only $\Delta_{\rm MS}$ bins with at least 6 galaxies are considered. The median stellar mass of the galaxy sample is indicated at the top of each redshift column. Fractions are corrected for statistical uncertainties as described in Section~\ref{sec:Corrections}. We find a rapid increase of the merger fraction that star formation episodes moving galaxies into the SB regime are almost always associated with a major merger.}
\end{figure*}

\begin{figure*}
\begin{center}
\includegraphics[width=0.45\textwidth,angle=90]{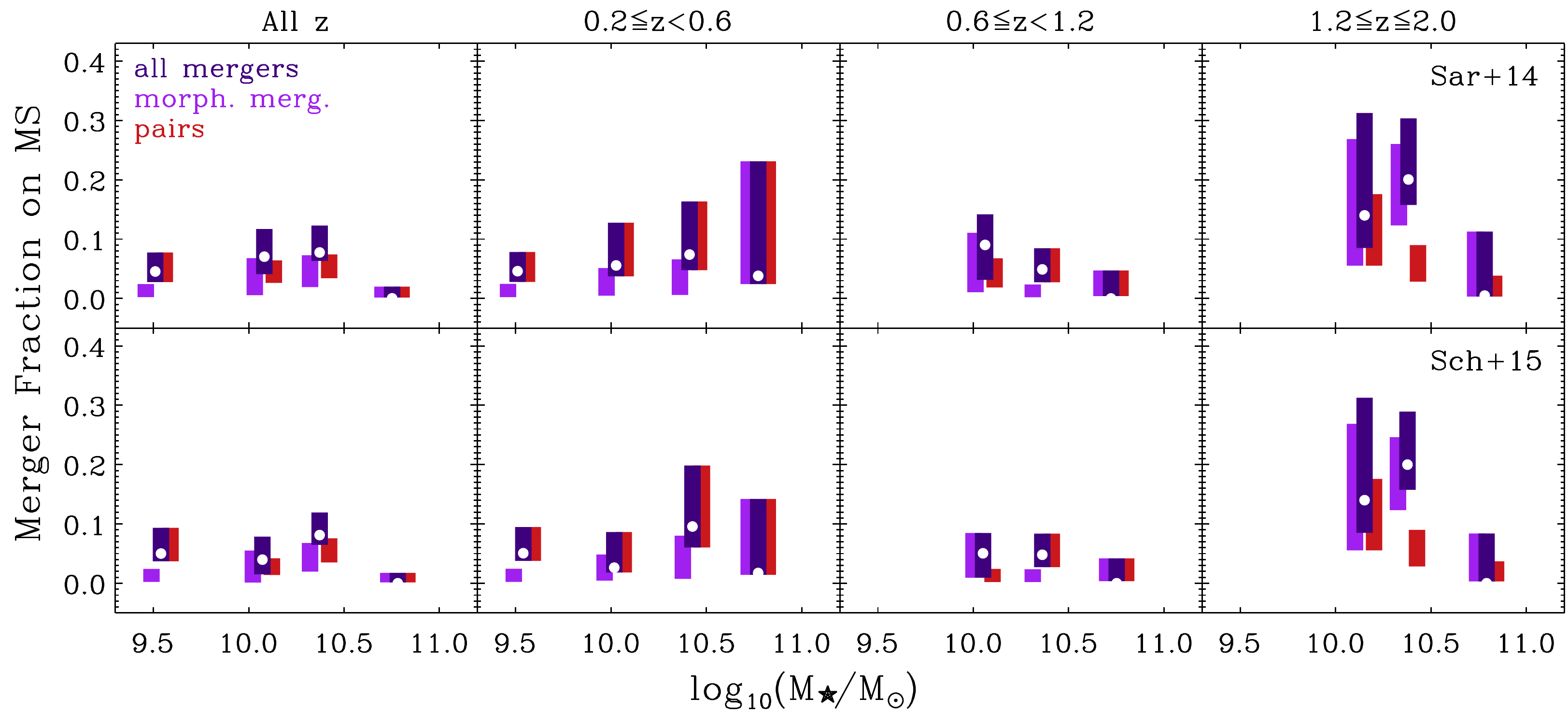}
\end{center}
\caption{\label{fig:FracMergerMass} Mass dependence of the merger fraction for galaxies that are lying on the MS, i.e., that have an offset from the MS locus within the range $-0.3 \leq \Delta_{\rm MS}\leq 0.3$. Symbols and colour scheme as in Fig.~\ref{fig:FracMerger}.
MS major mergers occur with roughly equal frequency at all masses ${\gtrsim}10^{10}\Msol$ in our sample}
\end{figure*}

We showed above that the global merger fraction is roughly constant within the MS. It is, however, interesting to check whether the MS merger fraction displays any dependence on galaxy stellar mass. In Fig.~\ref{fig:FracMergerMass} we show how the merger fraction varies with stellar mass for galaxies lying within the scatter of the MS, namely have $-0.3 \leq \Delta_{\rm MS}\leq 0.3$. As in Fig.~\ref{fig:FracMerger}, we also show results separately for morphological mergers and spectro-photometric pairs.

We find no clear trend with stellar mass, with an overall merger fraction that stays constant at $\sim$5--10\% over the $\sim$2\,dex mass range explored here. There is a tendency -- most clearly seen when galaxies from all redshifts are grouped (first column in Fig.~\ref{fig:FracMergerMass}) -- for the merger fraction to decrease at the highest masses ($M_{\star}$>$10^{10.5}\Msol$). The decrease is, however, not statistically significant in the individual redshift bins and should be confirmed on larger data sets. Similar results are found when considering morphological mergers or pairs only.

We can further explore the influence of stellar mass by investigating whether the average offset from the MS of merging galaxies, i.e. the merger-driven SFR enhancement, shows any dependence on the galaxy mass. 
Where our sample is complete at all redshifts (i.e. for galaxies with $\Delta_{\rm MS}{>}$-0.4\,dex  and $M{>}10^{10.2}\Msol$), we find a median SFR offset of mergers (morphologically identified and pairs) which is constant with mass and has a value of $\Delta_{\rm MS}{\sim}$0.2\,dex (in agreement with what already discussed in Section~\ref{sec:FracMerg} and Fig.~\ref{fig:HistDMS}).
Other theoretical and observational studies have investigated this issue, with contrasting results. 
Consistent with our findings, the observational study of local mergers by \citet{Carpineti+15} reports an almost mass-independent SFR offset between mergers and normal galaxies.
The recent numerical work of \citet{Martin+17} indicates instead that in the Horizon-AGN hydrodynamical simulation \citep{Dubois+14} the intensity of the merger-induced enhancement of star formation decreases with stellar mass, with differences between mergers and non-interacting galaxies being strongest at low redshift (z\,$<$\,0.3) and for $M_{\star}<10^{10.5}\Msol$.
Given that our sample, where complete, mostly probes the relatively high-mass end of the galaxy distribution and the $z\sim1$ regime, i.e. the parameter space with the smallest mass dependence in the simulations, these numerical predictions are a priori not inconsistent with our results\footnote{The distribution of $\Delta_{\rm MS}$ for mergers having $M_{\star}{<}10^{10}\Msol$ could suggest the emergence of a mass dependence at low stellar masses (see Fig.~\ref{fig:DeltaMS}), but incompleteness is affecting this region of the parameter space in our sample and no conclusion can therefore be drawn.}.

%=============================%
%==                FIGURE 8                    ==%
%=============================%
%=============================%

 \subsubsection{Redshift evolution}
An in-depth discussion of the evolution of the global merger fraction with redshift is beyond the scope of this paper; this would require a different sample selection to ensure homogenous sampling of the entire galaxy population -- i.e. including both quiescent and star-forming galaxies -- across redshift and stellar mass, whereas our sample largely consists of star-forming galaxies by construction.
We can however investigate how the merger fraction evolves for galaxies on the MS, which is well probed by our selection.
We show this in Fig.~\ref{fig:FracMerger_zevol}, again considering only galaxies with $-0.3 \leq \Delta_{\rm MS}\leq 0.3$.
As discussed in Section~\ref{sec:SampleCompleteness}, the 24\,$\mu$m selection used to define the 3D-$HST$ MS sample imposes a minimum stellar mass that varies with redshift. We therefore include in this analysis galaxies with $\log(M_{\star}/\Msol{)>}10.2$ at any redshift, i.e., that are above the mass completeness of the highest redshift bin.

We observe an increase of the total merger fraction from about 5\% at $z<1$ to 10--15\% at $z>1$. A fit to our data with the parametric function $f(z)=f(0)\times(1+z)^m$ returns a formal index value of $m{=}2.9\pm0.5$. While the pair fraction remains roughly constant, morphologically identified mergers display a more pronounced redshift evolution and drive the overall increase in the $1.2\leq z \leq 2$ bin. 
Some previous merger fraction measurements, either based on pair statistics \citep{Williams+11,Newman+12,Man+16,Mantha+18} or morphological classification \citep{Lotz+08,Conselice+09,Bluck+12}, are also presented in Fig.~\ref{fig:FracMerger_zevol} for reference. 
Overall, our merger fraction is in good agreement with previous findings. We stress however that we do not aim to perform a fully consistent comparison, as it would require a careful homogenization of all different samples in terms of  stellar mass, galaxy separation, merger mass ratio, etc., which we have not attempted (although where possible we selected studies that satisfy selection criteria similar to ours). The literature data are provided to enable the reader to place our results in the context of these previous findings, but the caveats preventing a one-to-one comparison should be kept in mind.

%%%%%%%%%%%%%%%%%%%%%%%%%%%%%%%%
%%%%%%%%        SEC.  DISCUSSION       %%%%%%%%%
%%%%%%%%%%%%%%%%%%%%%%%%%%%%%%%%
%=============================%
%==                FIGURE 9                    ==%
%=============================%
%=============================%

\section{Discussion}\label{sec:Discussion}

\subsection{Starburst--merger connection}

Combining our new approach for the morphological classification of mergers with a pair statistic, we found an increase in the merger fraction with distance from the MS.
These results are summarised for the entire galaxy sample in Fig.~\ref{fig:2SFComparison}, where we compare our merger factions as a function of $\Delta _{\rm MS}$  with literature estimates/predictions.

Our results confirm that even at $z$\,$\gtrsim$\,1 a large fraction, if not the totality, of galaxies classified as SB ($\Delta_{\rm MS}{\geq}0.6$) are undergoing a major merger, consistent with findings from previous high-$z$ morphological analyses \citep[e.g.][]{Kartaltepe+12,Hung+13} and recent interstellar medium studies \citep{Puglisi+17,Elbaz+17,Calabro+18}. 

In the SB regime, the merger fraction is in fact $\sim$70\% or higher depending on the assumptions made.
Mergers among SB galaxies are dominated by late-stage, disturbed systems, with a ratio of morphologically classified mergers relative to galaxy pairs of about 8:1.
This preponderance of strongly perturbed galaxies is not unexpected and supports a scenario in which strong SFR enhancements with respect to the MS (factors $\sim$6-10 or above) are generated only during the late stages of mergers (e.g. \citealt{Sanders_Mirabel_1996,Murphy+1996,Veilleux+02,Haan+11} for observational studies, \citealt{Mihos_Hernquist_94,Springel2000,Cox+06,DiMatteo+08,Bournaud+11,Fensch+17} for simulation-based results), while earlier merger stages induce only mild SFR variations, as observed in studies of close pairs at low and high redshift \citep[e.g.][]{Lin+07,Robaina+09,Hwang+11,Xu+12,Scudder+12,Pipino+14}.
%=============================%
%==                FIGURE 10                    ==%
%=============================%

\begin{figure}
\begin{center}
\includegraphics[width=0.47\textwidth]{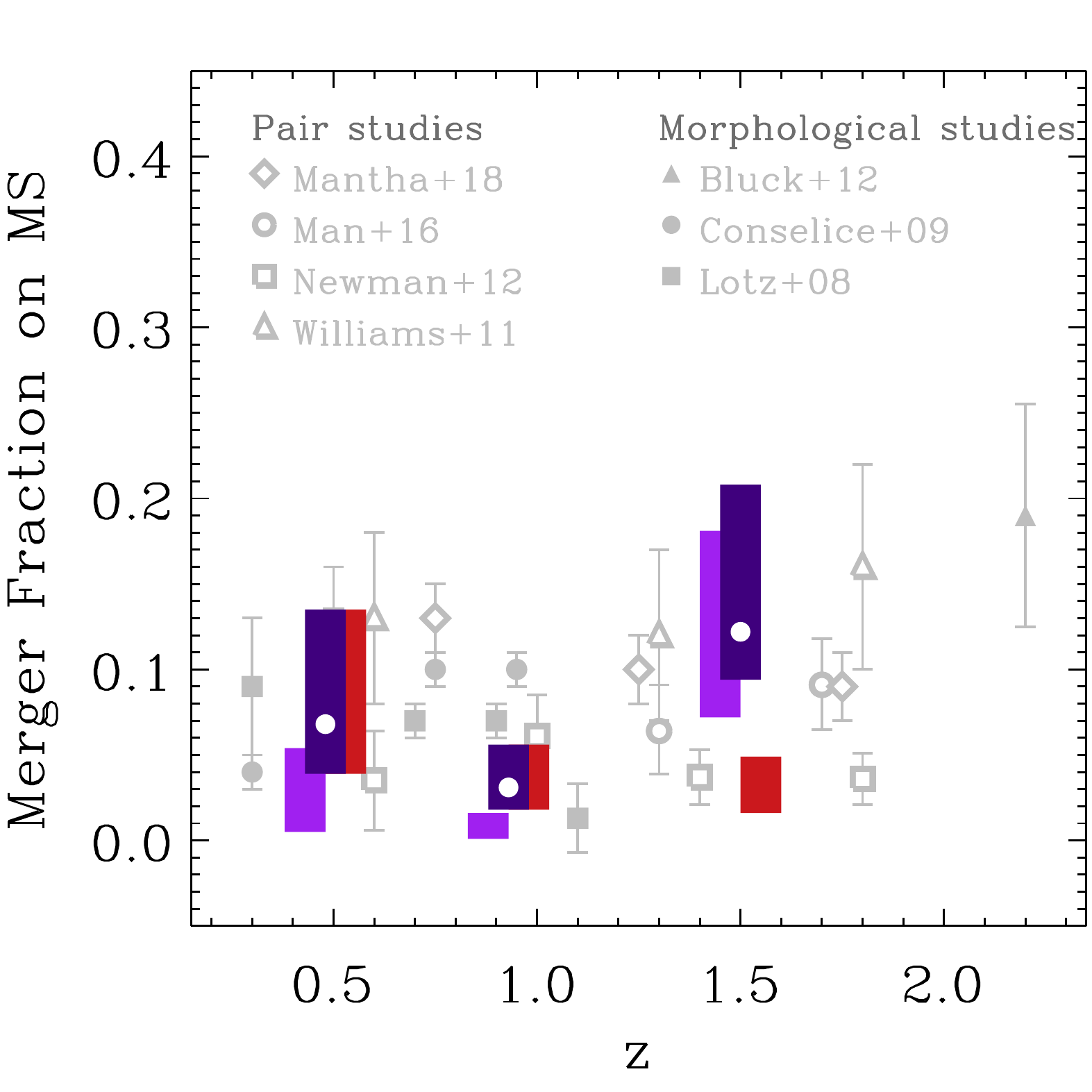}
\end{center}
\caption{\label{fig:FracMerger_zevol}Evolution of the merger fraction for galaxies that are on the MS. As in Fig.~\ref{fig:FracMerger}, coloured bars represent the global (dark purple), pair (red) and morphological (light purple) merger fraction measured for our galaxy sample. Only galaxies with $-0.3 \leq \Delta_{\rm MS}\leq 0.3$ and $\log(M_{\star}/\Msol{)>}10.2$ are here considered (only the \citealt{Sargent+14} MS definition is shown, but very similar results are obtained for the \citealt{Schreiber+15} parametrization). A compilation of literature measurements of the merger fraction from either close pairs \citep{Williams+11,Newman+12,Man+16,Mantha+18} or morphological studies \citep{Lotz+08,Conselice+09,Bluck+12} is shown with small grey symbols (see the figure legend). To match our MS selection, literature data are shown for star-forming galaxies only where possible \citep{Williams+11,Newman+12}. 
The MS total merger fraction increases from about 5\% at $z<1$ to 10-15\% at $z>1$, with this evolution being largely driven by morphologically identified mergers.}
\end{figure}
%=============================%
As a consequence of this, it is not surprising that mergers are also found within the MS, although they constitute a minority of this population. 
The existence of mergers on the MS is in agreement with previous findings \citep{Kartaltepe+12,Hung+13,Puech+14,Lackner+14,Wisnioski+15}, but we find a lower fraction than other studies at $z{\simeq}1$. In Fig.~\ref{fig:2SFComparison}, we plot with green triangles measurements by  \citet{Hung+13} and \citet{Kartaltepe+12} for comparison.
We perform this comparison at $0.5{\leq}z{\leq}{1.5}$, encompassing the highest redshift bins explored by these authors, in order to focus on the intermediate/high-$z$ regime where discs with clumpy substructure become frequent.
On the MS, we measure an overall merger fraction of $\sim$5--10\% at $0.5{\leq}z\leq{1.5}$.
In \citet{Hung+13}, about 20-30\% of MS galaxies are found to be interacting at a redshift 0.5${<}z{<}$1.5, while \citet{Kartaltepe+12} find a merger fraction exceeding 20\% for MS-like IR luminosities  at $z{=}1$ (see their fig. 9).
At $z{\simeq}2$ \citet{Kartaltepe+12} report a merger fraction of $\sim$24\% in ULIRGS on the MS. This is consistent within the errors  with our measured mergers fraction at 1.2${\leq}z{\leq}$2 (see rightmost panel in Fig.~\ref{fig:FracMerger}), although their estimate lies at the high end of our range.
As we discussed throughout the paper, our morphological merger selection is designed to minimise the misclassification of clumpy galaxies.  
The observed difference might therefore arise from a lower contamination of non-interacting galaxies in our merger sample, but different selections could also be responsible.
%=============================%
%==                FIGURE 11                  ==%
%=============================%
\begin{figure*}
\begin{center}
\includegraphics[width=0.99\textwidth]{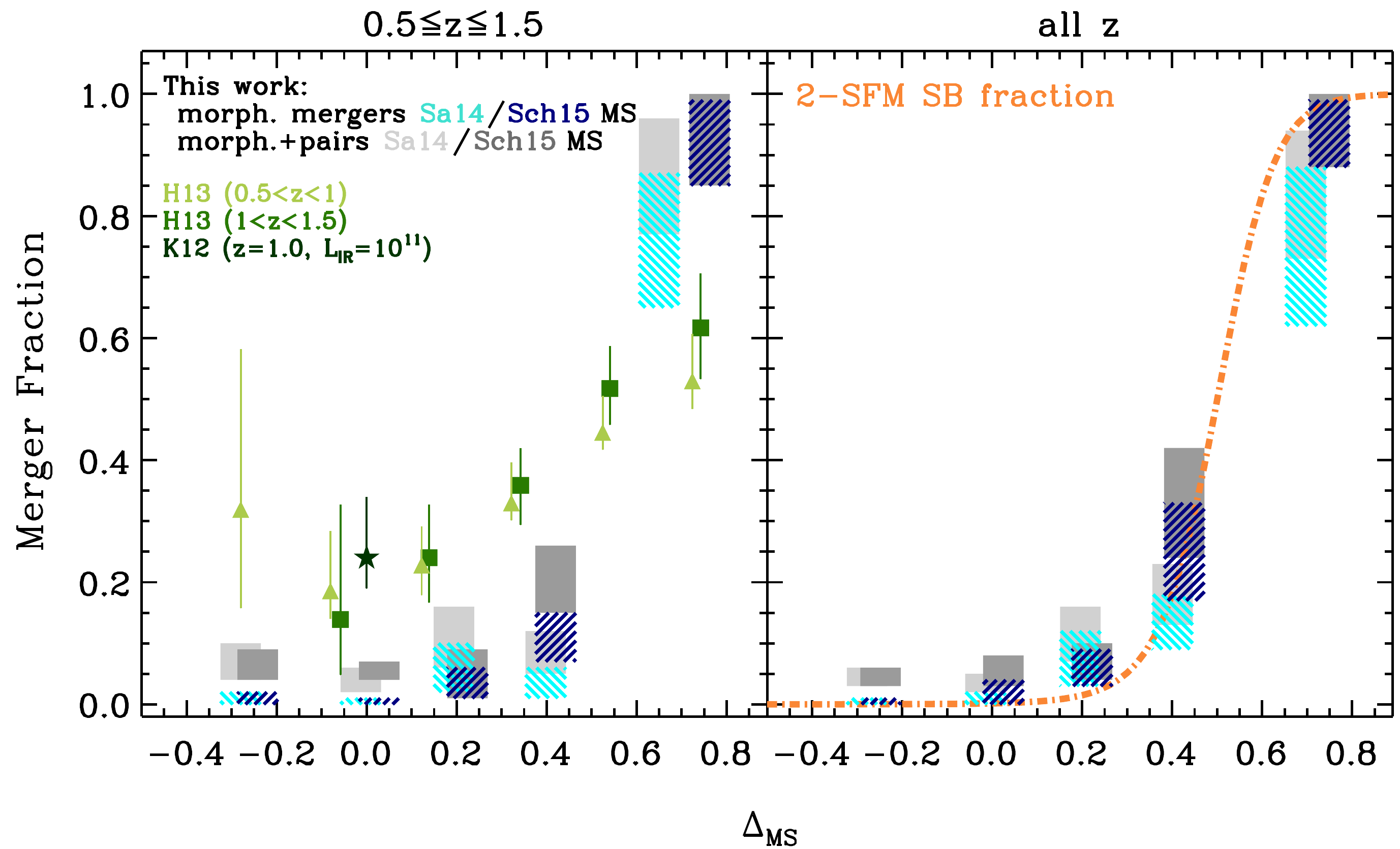}
\end{center}
\caption{\label{fig:2SFComparison} Comparison of our measured merger fraction with literature data. The left- and right-hand panels show the fraction as function of the distance from the MS for galaxies at intermediate redshift, $0.5{\leq}z{\leq}1.5$, and for galaxies at all redshifts, respectively.  
Dashed bars are the morphological merger fraction and filled grey bars in the background represent the total merger fraction (pairs+morphological mergers). 
Cyan and light grey correspond to the MS definition in \citet{Sargent+14}; and 
 blue and dark grey to the \citet{Schreiber+15} MS parametrization.
Light green triangles and dark green squares show the observed merger fractions in the IR-selected sample of \citet{Hung+13}. The dark star with error bar is the merger fraction measured in $z{=}1$ galaxies with $L_{\rm IR}=10^{11}\Lsol$ by \citet{Kartaltepe+12}.  The orange line on the right is the 2-SFM SB fraction from \citet{Sargent+12,Sargent+14}. 
We find a good agreement between our measured merger fraction and the 2-SFM SB prediction, especially if considering morphological mergers. This suggest a link between SB activity and final merger stages, both for high-sSFR SBs and for more moderate SFR enhancements that do not moves galaxies above the MS.}
\end{figure*}
%=============================%
Merging galaxies on the MS may be observed not only because early merger stages are associated with moderate SFR enhancements, but also as a consequence of high-$z$, gas-rich mergers being less efficient in triggering starbursts in general.
In hydrodynamical simulations of high-$z$ galaxies, even late-stage mergers induce only moderate SFR enhancements (factors of about 3-5)  due to the presence of an already highly turbulent medium prior to the interaction (\citealt{DiMatteo+08,Perret+14,Fensch+17}; specifics of the gas density distribution may also play a role, see \citealt{Sparre_Springel_2016}).
Recent ALMA observations have revealed the existence of a $z{\sim}2$ population of galaxies with compact, nuclear star formation and SB-like depletion times, but SFRs within the upper part of the MS \citep{Elbaz+17}. These SB galaxies hidden in the MS also display spheroid-like morphologies. The mergers on the MS found in our analysis could represent the preceding stage to this evolutionary phase leading to spheroid formation.
We note that most of our MS mergers are at lower masses than the mass range $M_{\star}>10^{11} \Msol$ explored in \citet{Elbaz+17},  small number statistics at these high masses may however play a role. 
Very late merger stages, associated with an already compact remnant might also be missed by a pure morphology selection and a combination of morphological criteria and information on burstiness \citep{Elbaz+17} or obscuration \citep[e.g.][]{Calabro+18} could ensure a full accounting of all mergers stages.

Finally, we compare our results on the merger fraction with the empirical ``2-Star Formation Mode" (2-SFM) framework of \citet{Sargent+12,Sargent+14}.
In the 2-SFM, SB galaxies are described by a ``boost factor" -- i.e., the SFR enhancement with respect to the pre-burst phase -- which is a continuous and increasing function of the galaxy position with respect to the MS. While representing a minority, in this picture SB galaxies exist on the MS and correspond to galaxies having experienced a moderate SFR boosting from an initial SFR that was close to or even below the MS (see fig. 6 in \citealt{Sargent+14}). 
We explore how well our measured major merger fractions agree with the continuous boost factor formalism in the 2-SFM by showing with the orange, dotted line in Fig.~\ref{fig:2SFComparison} the 2-SFM prediction of how the SB fraction varies with $\Delta_{\rm MS}$.

We find a good agreement between our measured global merger fraction (morphological mergers+pairs) and the 2-SFM SB fraction, which is roughly constant on the MS and rapidly increases toward the MS upper edge at $\Delta_{\rm MS}$\,$\simeq$\,0.5.
Interestingly, the low SB fraction on the MS in the 2-SFM matches particularly well with the fraction of morphological mergers. 
We interpret this as evidence that starburst activity is linked to the final merger phases (selected by our morphological criterion), for ``canonical" high-sSFR SB, as well as more moderate SB events that do not cause the host galaxies to rise above the MS.
In this picture, the higher observed \emph{total} merger fraction ($\sim$5\%) with respect to the 2-SFM SB fraction on the MS ($\sim$1\%) arises from pairs where the galaxy--galaxy interaction has not yet induced significant SFR variations.
 
\subsection{Merger time-scales}\label{sec:MergerTimes}

We can also use our measurement of the fraction of galaxies in early merger stages (i.e., in pairs) and in late stages (morphologically identified mergers) to attempt a derivation of the relative observability time-scales of these two phases.
In doing so, we assume that all galaxies in pairs will eventually merge (i.e. we neglect fly-by interactions) and experience a period during which they display a disturbed appearance, so that the ratio of the observed fractions, $f_{\rm morph}/f_{\rm pair}$, can be interpreted in a statistical sense as a proxy for the ratio of the corresponding time-scales, $\tau_{\rm morph}/\tau_{\rm obs, pair}$.

Fig.~\ref{fig:FracMerger_zevol} suggests a different evolution of the morphological and pair fractions, with the former increasing with redshift and the latter remaining almost constant or slightly decreasing.
Given the relatively small cosmic volume probed by the GOODS-N field, observational effects such as cosmic variance and Poisson uncertainties could be at the origin of this trend (e.g., \citealt{Man+16} find a lower major merger fraction at $z{>}1$ in GOODS-N with respect to  other 3D-$HST$ fields).
However, the observed difference could also be explained by a differential evolution of the observability time-scales of early- and late-stage mergers.

The time interval over which a galaxy displays morphological disturbances, and therefore is morphologically identifiable as a merger, is known to depend on the galaxy gas fraction ($f_{\rm gas}$), with gas-rich galaxies developing strong asymmetries for longer time periods than gas-poor ones \citep{Lotz+10}. Given that the gas fraction of galaxies increases with redshift \citep[e.g.,][]{Daddi+10,Geach+11,Bethermin+15,Genzel+15,Schinnerer+16,Tacconi+18}, morphologically disturbed mergers are plausibly visible for a longer time interval at high redshift. 
Although typically assumed to be constant, also the galaxy pair visibility may display a redshift dependence. In the recent work of \citet{Snyder+17} a pair observability time-scale varying as $\tau_{\rm pair}{\propto}(1+z)^{-2}$ is required to recover the intrinsic merger rate from the Illustris simulation \citep{Vogelsberger+14,Genel+14}.
%=============================%
%==                FIGURE 12                  ==%
%=============================%
\begin{figure}
\begin{center}
\includegraphics[width=0.47\textwidth]{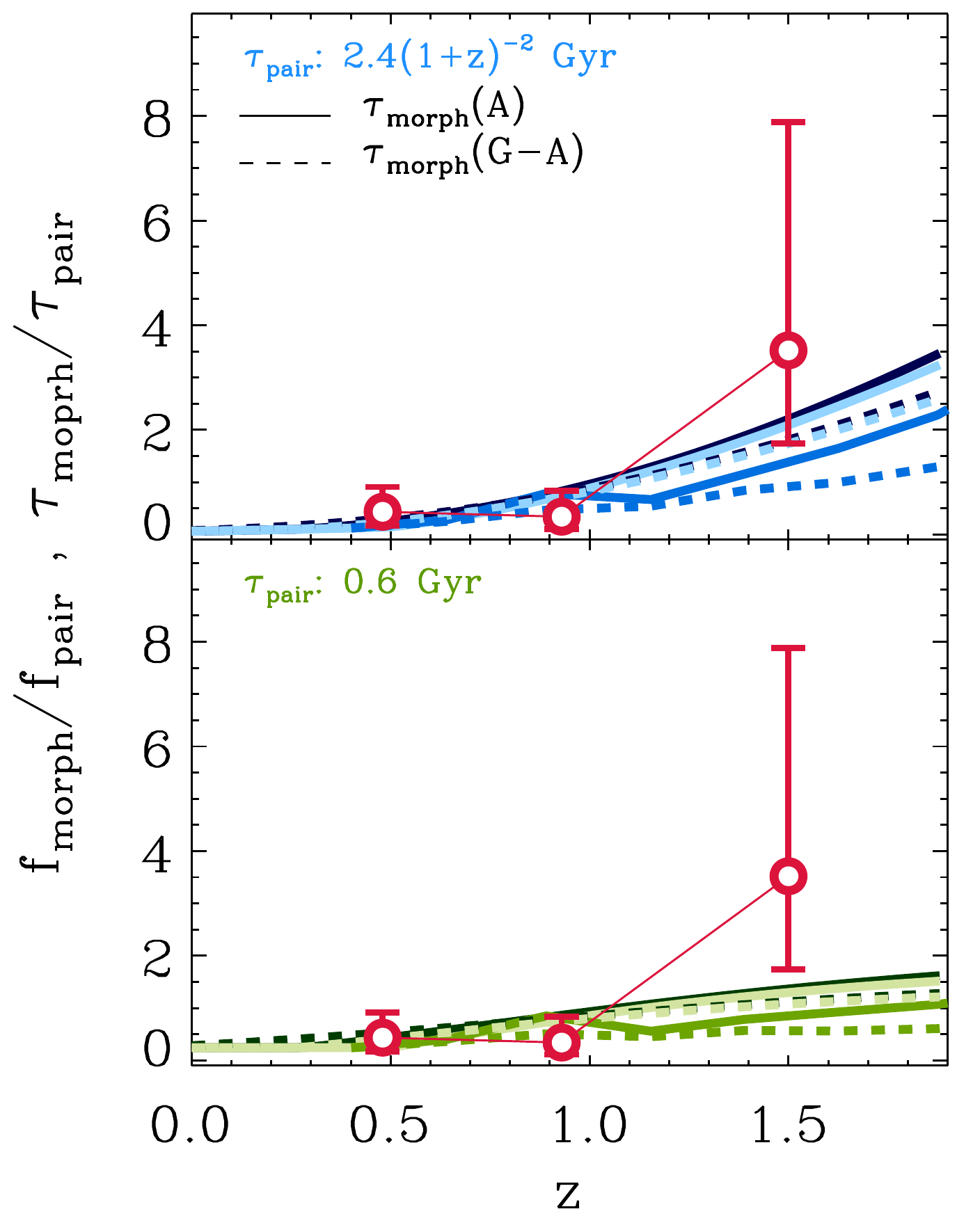}
\end{center}
\caption{\label{fig:time-scales} Ratio of the morphological merger and spectro-photometric pair fractions for MS galaxies versus redshift (red symbols with error bars; inferred from Fig.~\ref{fig:FracMerger_zevol}). Lines are predictions for the redshift evolution of the ratio between the visibility time-scale for mergers selected through morphological disturbances and for galaxy pairs, $\tau_{\rm morph}/\tau_{\rm pair}$. We derive $\tau_{\rm morph}(z)$ by combining the prescription of \citet{Lotz+10} for the dependence of $\tau_{\rm morph}$ on $f_{\rm gas}$ with literature measurements of gas fraction evolution for MS galaxies (dark shading -- 2-SFM, \citealt{Sargent+14}; medium shading -- \citealt{Bethermin+15}; light shading -- \citealt{Tacconi+18}. Gas fractions are estimated for a galaxy stellar mass $\log(M_{\star}/\Msol)=10.5$ and $\Delta_{\rm MS}=0$).
Solid and dashed lines show results for $\tau_{\rm morph}$ for mergers selected with the Asymmetry or Gini-Asymmetry method in \citet{Lotz+10}, respectively. For the upper panel, we adopt a redshift-dependent visibility time-scale for pairs, $\tau_{\rm pair}=2.4(1+z)^{-2}$\,Gyr, as measured by Illustris \citep{Snyder+17}. The lower panel assumes a constant $\tau_{\rm pair}=0.6$\,Gyr.
The differential redshift evolution of pairs and morphologically classified mergers we observe could be explained by an evolving observability time-scale for both pairs and morphological disturbances, as in the upper panel.
 }
\end{figure}
%=============================%

We test whether the ratio between our measured morphological merger and pair fractions is consistent with these scenarios in Fig.~\ref{fig:time-scales}, where we compare $f_{\rm morph}/f_{\rm pair}$ with the expected evolution of $\tau_{\rm morph}/\tau_{\rm pair}$ from numerical simulations.  
For  $\tau_{\rm pair}$, we assume either the aforementioned  \citet{Snyder+17} relation, $\tau_{\rm pair}=2.4(1+z)^{-2}$\,Gyr, or a constant value $\tau_{\rm pair}=0.6$\, Gyr, which is the average time-scale for pairs with separation $<$30 kpc and mass ratios between 1:1 and 4:1 in \citet{Lotz+10a,Lotz+11}. The evolution of the morphological observability time-scale is instead derived by combining the equations relating $\tau_{\rm morph}$ to $f_{\rm gas}$ in \citet{Lotz+10} (1:3 mass ratio) with recent literature estimates of the evolution of the gas fraction of MS galaxies \citep{Sargent+14,Bethermin+15,Tacconi+18}\footnote{The linear relations between $f_{\rm gas}$ - $\tau_{\rm morph}$ in \citet{Lotz+10} can formally lead to $\tau_{\rm morph}$ being equal or less than zero for very low gas fractions.
We therefore impose a minimal floor $T_0=0.15$\,Gyr to those relations, representing the minimum average observability times scale in the different models explored in \citealt{Lotz+11} (see their fig. 9), but our results hold even when this term is not included.}.
For our measured $f_{\rm morph}$ and $f_{\rm pair}$ we use the merger fractions derived for MS galaxies shown in Fig.~\ref{fig:FracMerger_zevol}. This ensures a consistent sampling of galaxy stellar mass and SFR --  which affect relevant galaxy properties like $f_{\rm gas}$ --  at all redshifts.
We also note that, although a minority in relative terms, mergers on the MS dominate the overall merger population in absolute numbers in our sample (see Fig.~\ref{fig:DeltaMS}). Due to the limited cosmic volume, the SB regime is sampled more heterogeneously at different redshifts, hampering evolutionary analysis for such galaxies due to low number statistics.
How the results presented here would change when considering the SB population only should be tested on larger data sets.

We caution that there are a number of aspects that make possible only an approximate comparison between our measured $f_{\rm morph}/f_{\rm pair}$ and the expected $\tau_{\rm morph}/\tau_{\rm pair}$: 
(i) \citet{Lotz+10} performed the morphological selection on $g-$band images rather than mass maps and they used a different combination of non-parametric structural estimators than ours. As a plausible range spanned by the simulations, we consider the $\tau_{\rm morph}$--$f_{\rm gas}$ relations derived for morphological criteria including the asymmetry parameter which is the most discriminating in our selection, but how these relations change if applied to the mass maps has not been tested\footnote{The MIRAGE simulations of \citet{Perret+14} which underpin our classification scheme cannot be used for this test as they all have a high initial gas fraction of 65\%.}. 
(ii) The simulated relations in \citet{Lotz+10} refer to total gas fractions, i.e. atomic plus molecular gas (HI+H$_2$). While at $z\gtrsim1$ the molecular phase presumably dominates the gas content, the contribution of HI becomes increasingly important at low redshift.
Observational constraints on the combined molecular and atomic content of galaxies are however not currently available out to high redshift and we thus rely on observational studies of the H$_2$ fraction for an estimate of $f_{\rm gas}$.  Actual $\tau_{\rm morph}$ values are likely higher than derived here for the local Universe. (iii) Finally, the details of the pair selection (primary mass, mass ratio and separation) in \citet{Snyder+17} slightly differ from those we employed and could have an impact on time-scale estimates.

Keeping these caveats in mind, Fig.~\ref{fig:time-scales} shows a reasonably good agreement between the evolution of the merger fraction ratio we measured here and the expected evolution of the ratio of time-scales. 
Although a constant $\tau_{\rm pair}$ cannot be ruled out, our $z{\sim}1.5$ measurement seems to favour a scenario in which the pair observability displays a dependence on redshift as found in \citet{Snyder+17}, with possibly an even a steeper evolution than the $\tau_{\rm pair}{\propto}(1+z)^{-2}$ proposed by these authors.
As specified earlier in this section, in our calculations we assumed that all galaxy pairs will eventually merge by $z$=0. In reality, some close pairs may never merge due to unbound systems and/or pairs falling in the long time-scale tail of the merger time-scale distribution.
If we assume that about 20-40\% (depending on exact pair selection criteria, e.g. \citealt{Kitzbichler_White_08,Kampczyk+13,Moreno+13}) of close pairs do not result in a merger, the correspondingly rescaled measurements of Fig.~\ref{fig:FracMerger_zevol} lead to a somewhat steeper redshift evolution in Fig.~\ref{fig:time-scales} (such that the highest redshift measurement becomes $\tau_{\rm morph}/\tau_{\rm pair}\,{\sim}$\,5). However these changes remain within the presently shown error bars.

%%%%%%%%%%%%%%%%%%%%%%%%%%%%%%%%
%%%%%%%%       SEC.  SUMMARY            %%%%%%%%%
%%%%%%%%%%%%%%%%%%%%%%%%%%%%%%%%

\section{Summary}\label{sec:Summary}

In this paper, we have studied the frequency of major mergers (down to a 1:6 mass ratio), selected either morphologically or through close pair statistics, in a sample of $0.2\leq z \leq 2$ star-forming galaxies in the GOODS-N field that have reliable SFR estimates from highly deblended FIR data.

A novelty of our study is that we perform the morphological classification of mergers on resolved stellar mass maps constructed from UV-to-NIR $HST$ imaging, rather than relying on single-band imaging as customarily done in studies of the merger fraction in the literature. This approach minimizes the contamination by clumpy, non-interacting galaxies at $z\gtrsim 1$ and is applied here to the study of the merger fraction for the first time.
We also derived statistical corrections for this mass-based morphological classification to recover intrinsic merger fractions, accounting for cross-contamination among the disc and merger population due to selection criteria and noise-related effects.

With the combined sample of both morphological mergers and spectro-photometric pairs we have studied how the  fraction of early- and late-stage mergers varies within and above the main sequence of star-forming galaxies.
Our findings can be summarised as follows:
\begin{enumerate}
\item At all redshifts $0.2{\leq}z{\leq}2.0$, virtually all galaxies displaying star formation enhancements typical of SB galaxies, i.e. $\Delta_{\rm MS}$\,$\geq$\,0.6, are undergoing a merger. The majority of these systems are morphologically disturbed, late-stage mergers, whereas galaxy pairs represent only a small fraction of galaxies at these high sSFRs. This paints a picture where galaxies move significantly above the MS as a consequence of merger-induced processes even at $z{>}1$, with strong starburst episodes developing primarily during coalescence or close to coalescence phases.
\item About 5\% to 10\% of MS galaxies also display morphological signatures of a late-stage merger or are found in a close pair system. The global merger fraction is roughly constant within the scatter of the MS, but it increases towards the upper edge before rapidly rising to almost unity in the SB region. Contrary to the SB regime, the MS hosts comparable fractions of pair galaxies and late stage mergers. 
On the MS we measure an overall lower merger fraction than reported in previous work. This is likely a consequence of the lower contamination from clumpy, non-interacting galaxies among merger candidates in our new classification scheme. 
\item Mergers on the MS seem to occur at all mass scales $M_{\star}{\gtrsim}10^{10}\Msol$, with the possible exception of a decrease at the highest masses $M_{\star}{\simeq}10^{11}\Msol$. 
\item The observed fraction of late-stage major mergers on and above the MS matches the distribution of starbursting galaxies in the empirical 2-SFM framework, where galaxies are subject to a continuously varying range of SFR boosts, depending on their MS offset. In this framework, SBs on the MS have experienced more moderate SFR enhancements (e.g. from a position in the lower half of the MS), while high-$\Delta_{\rm MS}$ outliers are associated with the strongest perturbations. The observed agreement reinforces the idea that these perturbations are linked to the late phases of galaxy mergers both on and above the MS. 
\item We find a different redshift evolution for the fractions of pairs and  morphological disturbed mergers on the MS, with the former staying roughly constant or slightly decreasing between $z{=}0$ and 2 and the latter instead roughly doubling. Although this trend should be confirmed with larger datasets, it could be explained by the combination of a pair visibility that \emph{decreases} with redshift, as suggested by recent numerical simulations, and a visibility window for morphological features that instead \emph{increases} as a consequence of higher gas fractions at $z{>}1$. 
\end{enumerate}

%%%%%%%%%%%%%%%%%%%%%%%%%%%%%%%%%%%%%%%%%%%%%%
%%%%%%%%%%%%%%%  ACKNOWLEDGMENTS    %%%%%%%%%%%%%%%%%%
%%%%%%%%%%%%%%%%%%%%%%%%%%%%%%%%%%%%%%%%%%%%%%

\section*{Acknowledgements}

We thank F. Owen for sharing the GOODS-N VLA 20\,cm radio maps used in Figure 1.
AC acknowledges support from the UK Science and Technology Facilities Council (STFC) consolidated grant ST/L000652/1. 
MTS acknowledges support from a Royal Society Leverhulme Trust Senior Research Fellowship (LT150041).
Support  for  this  work  was  provided  by  NASA through grant HST-GO-13872 from the Space Telescope Science Institute, which is operated by AURA, Inc., under NASA contract NAS 5-26555.  
PO further acknowledges support by the Swiss National Science Foundation.
This work is based on observations taken by the 3D-HST Treasury Program (GO 12177 and 12328) with the NASA/ESA HST, which is operated by the Association of Universities for Research in Astronomy, Inc., under NASA contract NAS5-26555.

%%%%%%%%%%%%%%%%%%%%%%%%%%%%%%%%%%%%%%%%%%%%%%
%%%%%%%%%%%%%%%%%%%% REFERENCES %%%%%%%%%%%%%%%%%%
%%%%%%%%%%%%%%%%%%%%%%%%%%%%%%%%%%%%%%%%%%%%%%

\bsp

%%%%%%%%%%%%%%%%%%%%%%%%%%%%%%%%%%%%%%%%%%%%%%
%%%%%%%%%%%%%%%%% APPENDICES %%%%%%%%%%%%%%%%%%%%%
%%%%%%%%%%%%%%%%%%%%%%%%%%%%%%%%%%%%%%%%%%%%%%

\appendix

\section{Parameters evolution in simulations and selection completeness}  \label{app:AsyCentroid}	
%=============================%
%==                FIGURE A1                  ==%
%=============================%
\begin{figure}
\begin{center}
\includegraphics[width=0.49\textwidth]{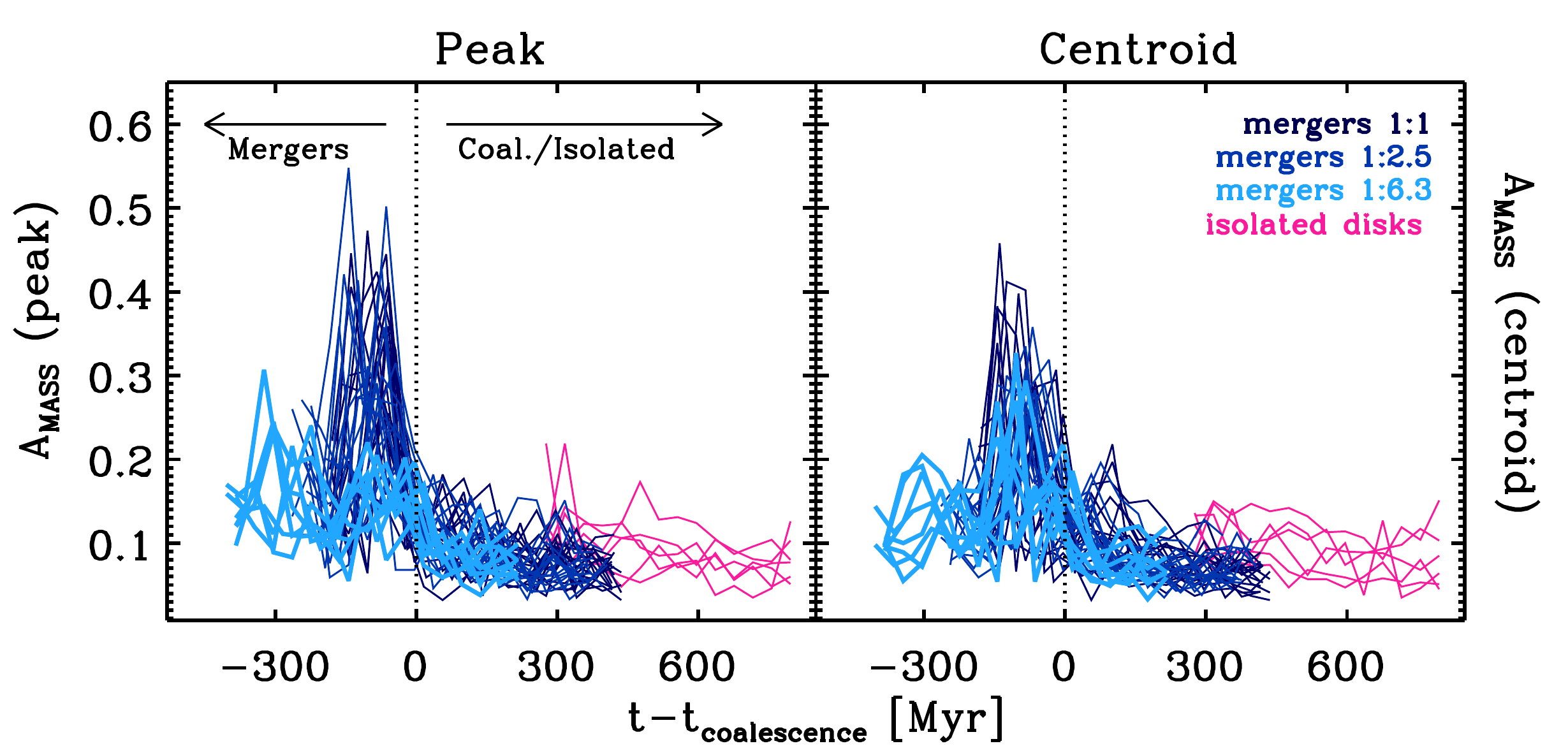}
\end{center}
\caption{\label{fig:AsyEvol} Evolution as a function of time (with respect to coalescence, thin dotted line a $t=0$) of the asymmetry index measured on the mass maps for the MIRAGE simulation of isolated discs (magenta lines) and mergers (blue). 
Darker shades of blue correspond to increasing merger ratio (see figure legend).
 The left-hand panel shows the evolution of the asymmetry index $A_{\star, \rm  peak}$,  calculated using the peak of the mass distribution as centre of rotation. In the right-hand panel is plotted the asymmetry calculated with respect to the centroid of mass, $A_{\star, \rm  centroid}$. For low-mass ratio mergers (1:6.3, light blue lines) the asymmetry variation during the merger phase is more clearly pronounced when using $A_{\star, \rm  centroid}$.}
\end{figure}
%=============================%

The asymmetry parameter is defined as the normalized residual flux (or mass in our case) that is measured in a galaxy image after subtraction of a 180$\degr$-rotated version of it. 
In the calculation of the asymmetry, a centre of rotation must therefore be chosen.
In Paper I we used for the merger classification solely the asymmetry derived using as centre of rotation the peak of the mass distribution, $A_{\star, \rm  peak}$, as it provides the clearest separation between mergers and discs (as also seen in Figure \ref{fig:Class}).
%=============================%
%==                FIGURE A2                  ==%
%=============================%
\begin{figure}
\begin{center}
\includegraphics[width=0.43\textwidth]{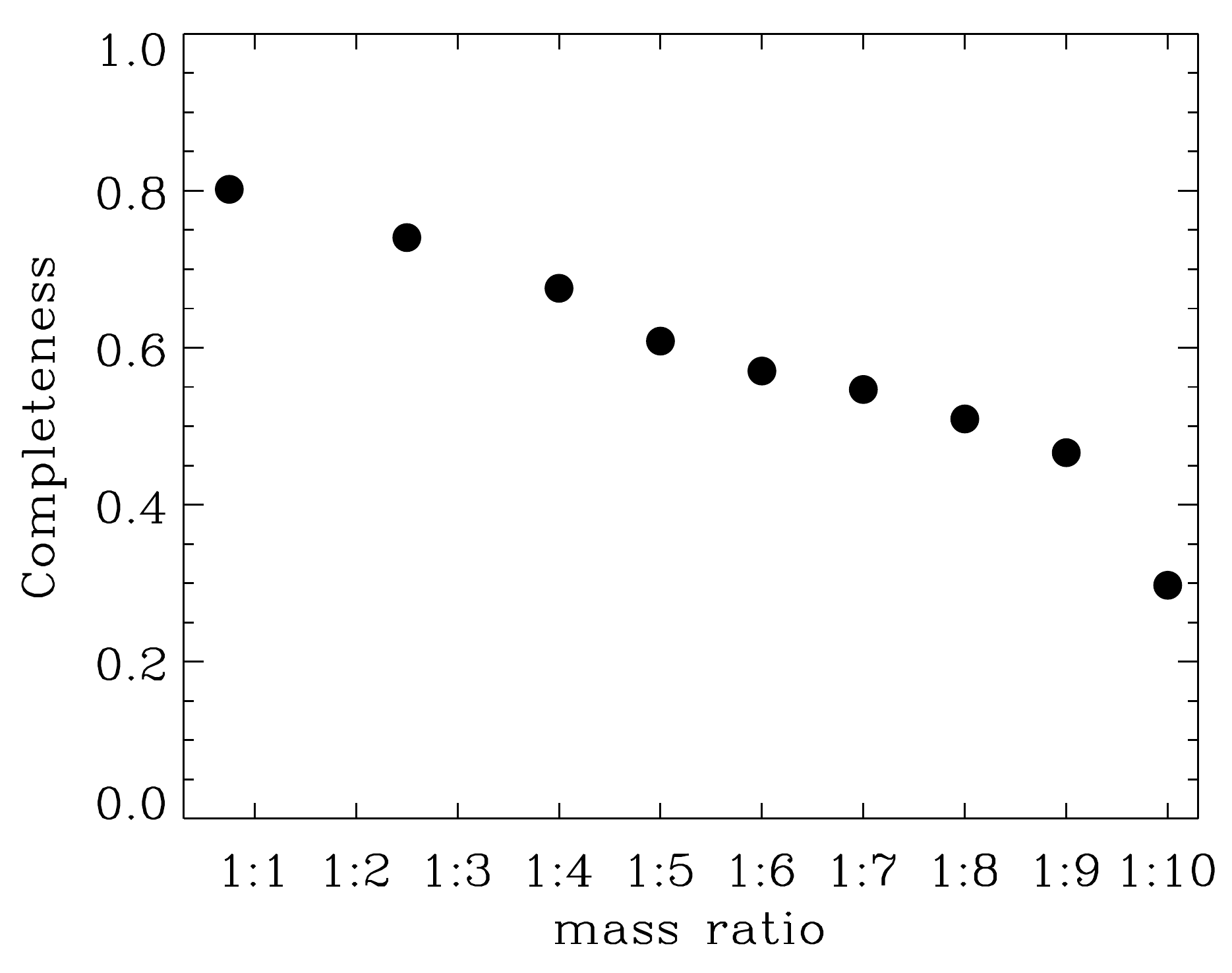}
\end{center}
\caption{\label{fig:Completeness} Completeness of the morphological merger selection as a function of the merger mass ratio.
The completeness is estimated on a set of $\sim$1000 artificial mergers generated from the superposition of the mass maps of random primary--secondary pairs selected from our sample. The galaxies in these artificial mergers were chosen to be at a similar redshift and to have a mass ratio in the range [1:1,1:10].}
\end{figure}
%=============================%
%=============================%
%==                FIGURE B1                 ==%
%=============================%
\begin{figure}
\begin{center}
\includegraphics[width=0.49\textwidth]{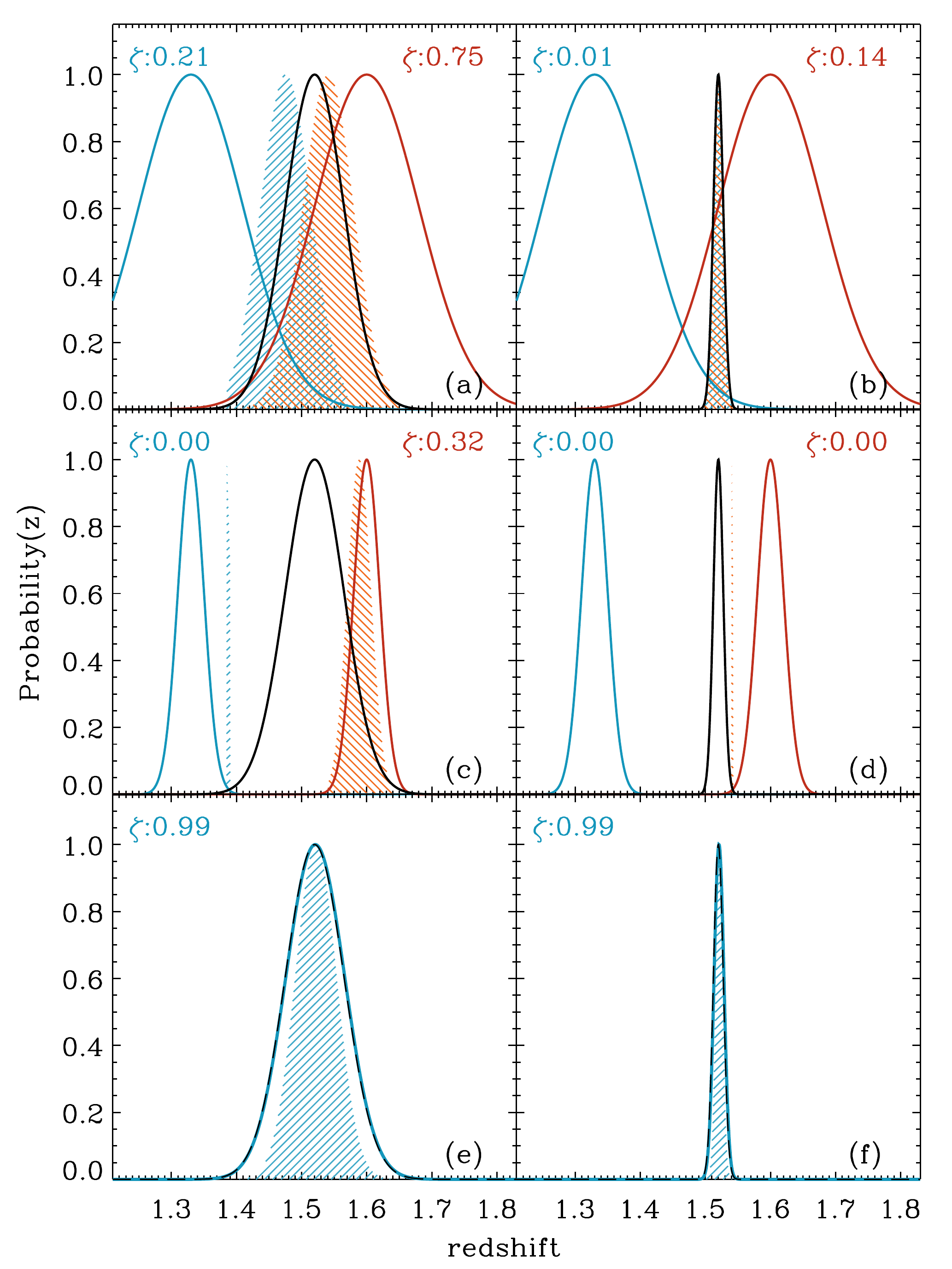}
\end{center}
\caption{\label{fig:PDFex} Illustration of the calculation of the pair probability distribution, $\zeta_{\rm pair}(z)$. The  probability distribution $P(z)$ of the primary galaxy is shown with a black solid Gaussian and is aways centred at $z_{\rm p}$\,=\,1.52. In panels \emph{a}, \emph{c} and \emph{e} the primary has $\sigma_{\rm p} =0.045$ (typical photo-$z$ error); in \emph{b}, \emph{d} and \emph{f}  it has $\sigma_{\rm p}=0.003(1+z)$\,=\, 0.0076 (typical spec-$z$ error).
The secondary galaxy can have $z_{\rm s}$\,=\,1.33 (blue Gaussian in panels \emph{a} to \emph{d}),  $z_{\rm s}$\,=\,1.60 (red) or identical redshift as the primary (panels \emph{e} and \emph{f}).
In panels panels \emph{a}  and \emph{b} the secondary galaxy has $\sigma_{\rm s}=0.08$, in \emph{c} and \emph{d}  it has $\sigma_{\rm s}=0.02$, and $\sigma_{\rm s} =\sigma_{\rm p}$ in panels \emph{e}  and \emph{f}. 
The resulting function $\zeta_{\rm pair}(z)$ is shown with the shaded areas matched in colour to the corresponding secondary galaxy. Note that, by definition, $\zeta_{\rm pair}(z) \neq 0$ only in the intersection range $[z_{\rm p}-3\sigma_{p}, z_{\rm p}+3\sigma_{p}]$\,$\cap$\,$[z_{\rm s}-3\sigma_{s}, z_{\rm s}+3\sigma_{s}]$.
Along the upper edge of each panel, the integral value of $\zeta_{\rm pair}(z)$ over this range is given, for each primary--secondary pair configuration.
  For plotting purposes, all probability distributions are normalized to the peak value.
}
\end{figure}
%=============================%

%=============================%
%==                FIGURE C1                 ==%
%=============================%
\begin{figure*}
\begin{center}
\includegraphics[width=0.45\textwidth,angle=90]{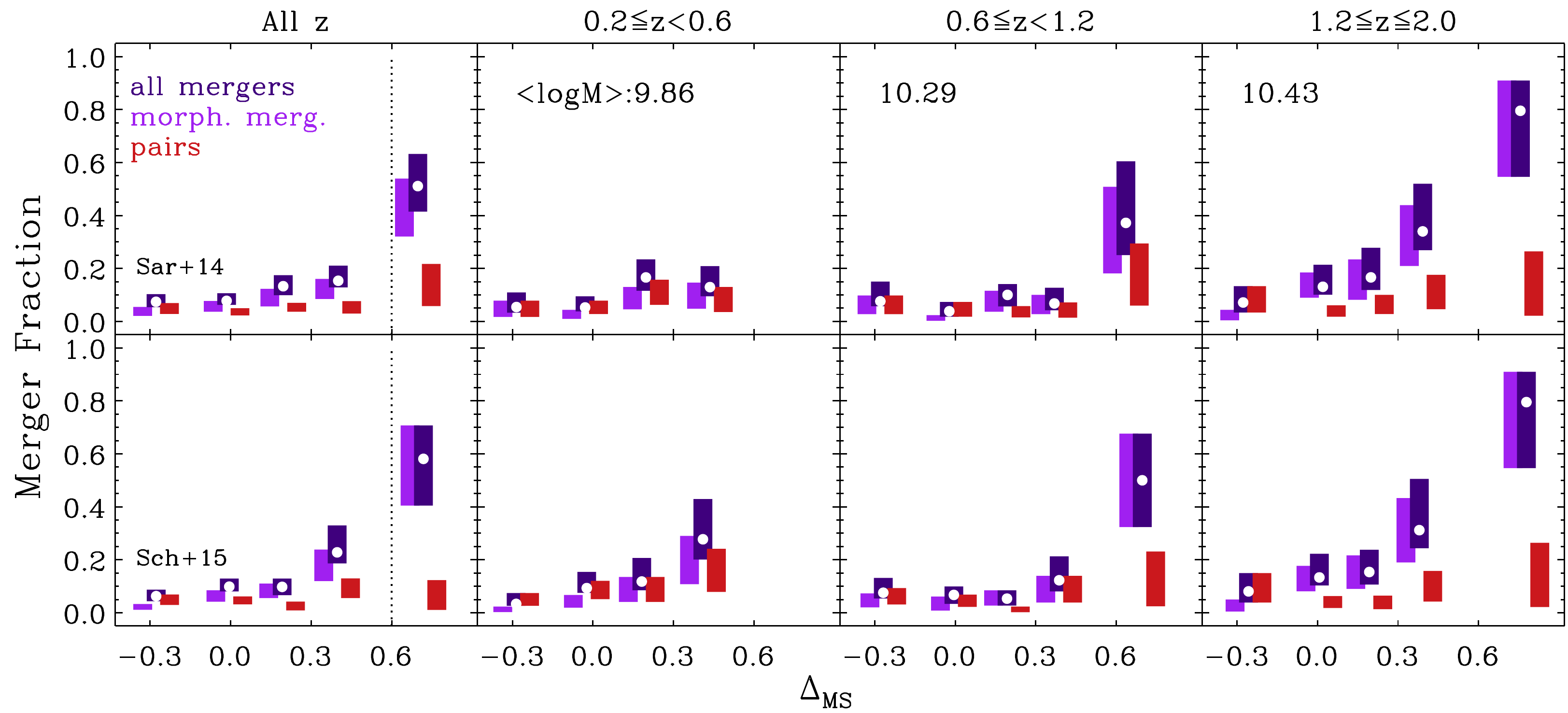}
\end{center}
\caption{\label{fig:FracMerger_noCorr} As in Fig.~\ref{fig:FracMerger} but we show here the merger fraction as directly calculated from the data, without applying any statistical correction to account for the completeness and contamination in the merger selection as described in Section~\ref{sec:Corrections}.}
\end{figure*}
%=============================%

It is however also possible to use the centroid of the mass distribution to define the asymmetry $A_{\star, \rm  centroid}$.
We show in Fig.~\ref{fig:AsyEvol} the evolution of the asymmetry index measured in these two different fashions for the MIRAGE simulated galaxies. 
In the simulations, mergers with lower mass ratios (1:6.3) display a clear increase in $A_{\star, \rm  centroid}$ during the merger phase, whereas a less marked asymmetry variation is shown when using $A_{\star, \rm  peak}$.
In combination with the $M_{20}$ index, the inclusion of this alternative asymmetry definition  therefore increases the completeness of the classification for minor mergers and for this reason was included in this study, as described in Section~\ref{sec:MorphoClass}.

To quantify the level of completeness of our selection criteria and how this depends on the mass ratio of the merging galaxies, we also generated artificial mergers by selecting a random subset of galaxies from our sample and by superimposing to each of their mass map the map from a secondary galaxy with a similar redshift ($|\Delta z| \le 0.25$).
We selected secondary galaxies with a mass ratio from 1:1 to 1:10 and placed it  at a random position with a spatial separation 1.5\,$<$r$<$10\,kpc at the galaxy redshift (therefore excluding the ``coalescence phase" and well separated pairs, which are not detected by the morphological technique).
The results are shown in Fig.~\ref{fig:Completeness}. 
The morphological selection provides a high completeness $\gtrsim 70$\% for major mergers down to a ratio 1:4, but decreases to $<$60\% for mass ratios smaller than 1:6.
In the 1:1 to 1:6 range, we estimate an average completeness of 70\%.

\section{Pair probability distribution}  \label{app:PhotometricPairs}	
In Fig.~\ref{fig:PDFex} we illustrate a range of possibilities for the pair probability distribution  $\zeta_{\rm pair}(z)$ function discussed in Section~\ref{sec:spec-photo_pairs}.
In all these examples the redshift distribution $P_{\rm p}(z)$ of the primary galaxy is centred at the same redshift ($z_{\rm p}$\,=\,1.52) and we set its standard deviation to be equal to either the typical photo-$z$ error in our sample, $\sigma_{\rm p} =0.045$ (left-hand side panels in Fig.~\ref{fig:PDFex}), or to the spec-$z$ error, $\sigma_{\rm p}=0.003(1+z_{\rm p})$\,=\, 0.0076 (right-hand side).
We consider different combinations with secondary galaxies that have different or identical redshift as the primary, as well as a broad or  narrow redshift distribution.
For each of these cases, we show the functional form of $\zeta_{\rm pair}(z)$ and provide the value of the integral of $\zeta_{\rm pair}(z)$, i.e., the fraction of the primary galaxy that is in a pair with the secondary.

As illustrated in Fig.~\ref{fig:PDFex} and expected, with this formalism a primary galaxy will provide a non-negligible contribution to the number of merging galaxies  only if the secondary $P_{\rm s}(z)$  has similar peak and standard deviation values.
This also implies that primary and secondary galaxies with identical $P(z)$ will result in a integrated $\zeta_{\rm pair}$\,$\simeq$\,1, regardless of the width of the Gaussian as shown in panels \emph{e} and \emph{f}.
Although this can be counter-intuitive, it can be understood as follows. In panel \emph{f} we can determine with a high precision that the two galaxies are a pair at the same redshift. In panel \emph{e}, on the other hand, the redshift measurement has a much lower precision. This means that we cannot pinpoint their location in redshift, but consequently we cannot firmly rule out that they are a pair either.
We also note that in the case of a very broad redshift distribution that spans multiple redshift bins, a primary galaxy will provide a small contribution to both the number of merging galaxies in the numerator of Equation~\ref{eq:f_merge} and to the total number of galaxies in the denominator.

For the sample considered in this study, the large majority of the galaxies with a companion  (70\%)  are in configuration similar to the one shown in panel (b), namely having a spectroscopic redshift for the primary and a photometric redshift for the neighbouring galaxy.

\section{Merger Fraction with no Corrections}\label{app:FracnoCorrection} 
For reference, we show in Fig.~\ref{fig:FracMerger_noCorr} the merger fractions as directly measured from the data, without applying the statistical corrections to the morphological selected mergers and spectro-photometric pairs described in Section~\ref{sec:Corrections}.

\label{lastpage}

\begin{thebibliography}{99}
\bibitem[Alexander et al.(2003)]{Alexander+03} Alexander, D.~M., Bauer, F.~E., Brandt, W.~N., et al.\ 2003, \aj, 126, 539 
\bibitem[Armus et al.(1987)]{Armus+87} Armus, L., Heckman, T., \& Miley, G.\ 1987, \aj, 94, 831 
\bibitem[Arnouts et al.(1999)]{Arnouts+99} Arnouts, S., Cristiani, S., Moscardini, L., et al.\ 1999, \mnras, 310, 540 
\bibitem[Arribas et al.(2004)]{Arribas+04} Arribas, S., Bushouse, H., Lucas, R.~A., Colina, L., \& Borne, K.~D.\ 2004, \aj, 127, 2522 
\bibitem[Barnes(1988)]{Barnes_1988} Barnes, J.~E.\ 1988, \apj, 331, 699
\bibitem[Bell et al.(2005)]{Bell+05} Bell, E.~F., Papovich, C., Wolf, C., et al.\ 2005, \apj, 625, 23  
\bibitem[B{\'e}thermin et al.(2011)]{Bethermin+11} B{\'e}thermin, M., Dole, H., Lagache, G., Le Borgne, D., \& Penin, A.\ 2011, \aap, 529, A4 
\bibitem[B{\'e}thermin et al.(2015)]{Bethermin+15} B{\'e}thermin, M., Daddi, E., Magdis, G., et al.\ 2015, \aap, 573, A113 
\bibitem[Bluck et al.(2012)]{Bluck+12} Bluck, A.~F.~L., Conselice, C.~J., Buitrago, F., et al.\ 2012, \apj, 747, 34 
\bibitem[Bouch{\'e} et al.(2010)]{Bouche+10} Bouch{\'e}, N., Dekel, A., Genzel, R., et al.\ 2010, \apj, 718, 1001 
\bibitem[Bournaud et al.(2007)]{Bournaud+07} Bournaud, F., Elmegreen, B.~G., \& Elmegreen, D.~M.\ 2007, \apj, 670, 237 
\bibitem[Bournaud et al.(2008)]{Bournaud+08} Bournaud, F., Daddi, E., Elmegreen, B.~G., et al.\ 2008, \aap, 486, 741
\bibitem[Bournaud et al.(2011)]{Bournaud+11} Bournaud, F., Chapon, D., Teyssier, R., et al.\ 2011, \apj, 730, 4  
\bibitem[Brammer et al.(2012)]{Brammer+12} Brammer, G.~B., van Dokkum, P.~G., Franx, M., et al.\ 2012, \apjs, 200, 13 
\bibitem[Brinchmann et al.(2004)]{Brinchmann+04} Brinchmann, J., Charlot, S., White, S.~D.~M., et al.\ 2004, \mnras, 351, 1151 
\bibitem[Bruzual \& Charlot(2003)]{Bruzual_Charlot_2003} Bruzual, G., \& Charlot, S.\ 2003, \mnras, 344, 1000
\bibitem[Bundy et al.(2009)]{Bundy+09} Bundy, K., Fukugita, M., Ellis, R.~S., et al.\ 2009, \apj, 697, 1369 
\bibitem[Calabr{\`o} et al.(2018)]{Calabro+18} Calabr{\`o}, A., Daddi, E., Cassata, P., et al.\ 2018, \apjl, 862, L22 
\bibitem[Calzetti et al.(2000)]{Calzetti+00} Calzetti, D., Armus, L., Bohlin, R.~C., et al.\ 2000, \apj, 533, 682
\bibitem[Carpineti et al.(2015)]{Carpineti+15} Carpineti, A., Kaviraj, S., Hyde, A.~K., et al.\ 2015, \aap, 577, A119  
\bibitem[Caputi et al.(2007)]{Caputi+07} Caputi, K.~I., Lagache, G., Yan, L., et al.\ 2007, \apj, 660, 97 
\bibitem[Casey et al.(2012)]{Casey+12} Casey, C.~M., Berta, S., B{\'e}thermin, M., et al.\ 2012, \apj, 761, 140 
\bibitem[Chabrier(2003)]{Chabrier_2003} Chabrier, G.\ 2003, \pasp, 115, 763 
\bibitem[Cibinel et al.(2015)]{Cibinel+15} Cibinel, A., Le Floc'h, E., Perret, V., et al.\ 2015 \apj, 805, 181 (Paper I) 
\bibitem[Clements et al.(1996)]{Clements+96} Clements, D.~L., Sutherland, W.~J., McMahon, R.~G., \& Saunders, W.\ 1996, \mnras, 279, 477 
\bibitem[Conselice(2003)]{Conselice_2003} Conselice, C.~J.\ 2003, \apjs, 147, 1
\bibitem[Conselice et al.(2009)]{Conselice+09} Conselice, C.~J., Yang, C., \& Bluck, A.~F.~L.\ 2009, \mnras, 394, 1956 
\bibitem[Cox et al.(2006)]{Cox+06} Cox, T.~J., Jonsson, P., Primack, J.~R., \& Somerville, R.~S.\ 2006, \mnras, 373, 1013 
\bibitem[Cowie et al.(1995)]{Cowie+95} Cowie, L.~L., Hu, E.~M., \& Songaila, A.\ 1995, \aj, 110, 1576 
\bibitem[Daddi et al.(2007)]{Daddi+07} Daddi, E., Dickinson, M., Morrison, G., et al.\ 2007, \apj, 670, 156 
\bibitem[Daddi et al.(2010b)]{Daddi+10} Daddi, E., Elbaz, D., Walter, F., et al.\ 2010, \apjl, 714, L118 
\bibitem[Di Matteo et al.(2005)]{DiMatteo+05} Di Matteo, T., Springel, V., \& Hernquist, L.\ 2005, \nat, 433, 604 
\bibitem[Di Matteo et al.(2008)]{DiMatteo+08} Di Matteo, P., Bournaud, F., Martig, M., et al.\ 2008, \aap, 492, 31 
\bibitem[Duc et al.(1997)]{Duc+97} Duc, P.-A., Mirabel, I.~F., \& Maza, J.\ 1997, \aaps, 124, 533 
\bibitem[Dubois et al.(2014)]{Dubois+14} Dubois, Y., Pichon, C., Welker, C., et al.\ 2014, \mnras, 444, 1453 
\bibitem[Elbaz et al.(2007)]{Elbaz+07} Elbaz, D., Daddi, E., Le Borgne, D., et al.\ 2007, \aap, 468, 33 
\bibitem[Elbaz et al.(2011)]{Elbaz+11} Elbaz, D., Dickinson, M., Hwang, H.~S., et al.\ 2011, \aap, 533, A119 
\bibitem[Elbaz et al.(2017)]{Elbaz+17} Elbaz, D., Leiton, R., Nagar, N., et al.\ 2017, arXiv:1711.10047 
\bibitem[Ellison et al.(2013)]{Ellison+13} Ellison, S.~L., Mendel, J.~T., Scudder, J.~M., Patton, D.~R., \& Palmer, M.~J.~D.\ 2013, \mnras, 430, 3128 
\bibitem[Elmegreen et al.(2007)]{Elmegreen+07} Elmegreen, D.~M., Elmegreen, B.~G., Ravindranath, S., \& Coe, D.~A.\ 2007, \apj, 658, 763 
\bibitem[Engelbracht et al.(1998)]{Engelbracht+98} Engelbracht, C.~W., Rieke, M.~J., Rieke, G.~H., Kelly, D.~M., \& Achtermann, J.~M.\ 1998, \apj, 505, 639 
\bibitem[Farrah et al.(2001)]{Farrah+01} Farrah, D., Rowan-Robinson, M., Oliver, S., et al.\ 2001, \mnras, 326, 1333 
\bibitem[Fensch et al.(2017)]{Fensch+17} Fensch, J., Renaud, F., Bournaud, F., et al.\ 2017, \mnras, 465, 1934 
\bibitem[F{\"o}rster Schreiber et al.(2009)]{Forster_Schreiber+09} F{\"o}rster Schreiber, N.~M., Genzel, R., Bouch{\'e}, N., et al.\ 2009, \apj, 706, 1364 
\bibitem[F{\"o}rster Schreiber et al.(2011)]{Forster_Schreiber+11} F{\"o}rster Schreiber, N.~M., Shapley, A.~E., Genzel, R., et al.\ 2011, \apj, 739, 45 
\bibitem[Giavalisco et al.(2004)]{Giavalisco+04} Giavalisco, M., Ferguson, H.~C., Koekemoer, A.~M., et al.\ 2004, \apjl, 600, L93
\bibitem[Geach et al.(2011)]{Geach+11} Geach, J.~E., Smail, I., Moran, S.~M., et al.\ 2011, \apjl, 730, L19 
\bibitem[Geach et al.(2017)]{Geach+17} Geach, J.~E., Dunlop, J.~S., Halpern, M., et al.\ 2017, \mnras, 465, 1789
\bibitem[Genel et al.(2014)]{Genel+14} Genel, S., Vogelsberger, M., Springel, V., et al.\ 2014, \mnras, 445, 175  
\bibitem[Genzel et al.(1998)]{Genzel+98} Genzel, R., Lutz, D., Sturm, E., et al.\ 1998, \apj, 498, 579 
\bibitem[Genzel et al.(2006)]{Genzel+06} Genzel, R., Tacconi, L.~J., Eisenhauer, F., et al.\ 2006, \nat, 442, 786 
\bibitem[Genzel et al.(2010)]{Genzel+10} Genzel, R., Tacconi, L.~J., Gracia-Carpio, J., et al.\ 2010, \mnras, 407, 2091 
\bibitem[Genzel et al.(2015)]{Genzel+15} Genzel, R., Tacconi, L.~J., Lutz, D., et al.\ 2015, \apj, 800, 20 
\bibitem[Gerhard(1981)]{Gerhard_1981} Gerhard, O.~E.\ 1981, \mnras, 197, 179 
\bibitem[Grogin et al.(2011)]{Grogin+11} Grogin, N.~A., Kocevski, D.~D., Faber, S.~M., et al.\ 2011, \apjs, 197, 35
\bibitem[Guo et al.(2012)]{Guo+12} Guo, Y., Giavalisco, M., Ferguson, H.~C., Cassata, P., \& Koekemoer, A.~M.\ 2012, \apj, 757, 120 
\bibitem[Guo et al.(2015)]{Guo+15} Guo, Y., Ferguson, H.~C., Bell, E.~F., et al.\ 2015, \apj, 800, 39 
\bibitem[Haan et al.(2011)]{Haan+11} Haan, S., Surace, J.~A., Armus, L., et al.\ 2011, \aj, 141, 100 
\bibitem[Hopkins et al.(2006)]{Hopkins+06} Hopkins, P.~F., Hernquist, L., Cox, T.~J., et al.\ 2006, \apjs, 163, 1 
\bibitem[Hopkins et al.(2008)]{Hopkins+08} Hopkins, P.~F., Cox, T.~J., \& Hernquist, L.\ 2008, \apj, 689, 17-48 
\bibitem[Hung et al.(2013)]{Hung+13} Hung, C.-L., Sanders, D.~B., Casey, C.~M., et al.\ 2013, \apj, 778, 129 
\bibitem[Hwang et al.(2011)]{Hwang+11} Hwang, H.~S., Elbaz, D., Dickinson, M., et al.\ 2011, \aap, 535, A60 
\bibitem[Ilbert et al.(2006)]{Ilbert+06} Ilbert, O., Arnouts, S., McCracken, H.~J., et al.\ 2006, \aap, 457, 841 
\bibitem[Jogee et al.(2009)]{Jogee+09} Jogee, S., Miller, S.~H., Penner, K., et al.\ 2009, \apj, 697, 1971 
\bibitem[Joseph \& Wright(1985)]{Joseph_Wright_1985} Joseph, R.~D., \& Wright, G.~S.\ 1985, \mnras, 214, 87 
\bibitem[Kampczyk et al.(2013)]{Kampczyk+13} Kampczyk, P., Lilly, S.~J., de Ravel, L., et al.\ 2013, \apj, 762, 43 
\bibitem[Karim et al.(2011)]{Karim+11} Karim, A., Schinnerer, E., Mart{\'{\i}}nez-Sansigre, A., et al.\ 2011, \apj, 730, 61 
\bibitem[Kartaltepe et al.(2007)]{Kartaltepe+07} Kartaltepe, J.~S., Sanders, D.~B., Scoville, N.~Z., et al.\ 2007, \apjs, 172, 320 
\bibitem[Kartaltepe et al.(2010)]{Kartaltepe+10} Kartaltepe, J.~S., Sanders, D.~B., Le Floc'h, E., et al.\ 2010, \apj, 721, 98 
\bibitem[Kartaltepe et al.(2012)]{Kartaltepe+12} Kartaltepe, J.~S., Dickinson, M., Alexander, D.~M., et al.\ 2012, \apj, 757, 23 
\bibitem[Kaviraj et al.(2013)]{Kaviraj+13} Kaviraj, S., Cohen, S., Windhorst, R.~A., et al.\ 2013, \mnras, 429, L40 
\bibitem[Kaviraj(2014)]{Kaviraj+14} Kaviraj, S.\ 2014, \mnras, 440, 2944 
\bibitem[Kennicutt(1998)]{Kennicutt_98} Kennicutt, R.~C., Jr.\ 1998, \araa, 36, 189 
\bibitem[Kitzbichler \& White(2008)]{Kitzbichler_White_08} Kitzbichler, M.~G., \& White, S.~D.~M.\ 2008, \mnras, 391, 1489 
\bibitem[Koekemoer et al.(2011)]{Koekemoer+11} Koekemoer, A.~M., Faber, S.~M., Ferguson, H.~C., et al.\ 2011, \apjs, 197, 36 
\bibitem[Lackner et al.(2014)]{Lackner+14} Lackner, C.~N., Silverman, J.~D., Salvato, M., et al.\ 2014, \aj, 148, 137 
\bibitem[Larson et al.(2016)]{Larson+16} Larson, K.~L., Sanders, D.~B., Barnes, J.~E., et al.\ 2016, \apj, 825, 128 
\bibitem[Law et al.(2007)]{Law+07} Law, D.~R., Steidel, C.~C., Erb, D.~K., et al.\ 2007, \apj, 656, 1 
\bibitem[Le Floc'h et al.(2005)]{LeFloch+05} Le Floc'h, E., Papovich, C., Dole, H., et al.\ 2005, \apj, 632, 169 
\bibitem[Leitner(2012)]{Leitner+12} Leitner, S.~N.\ 2012, \apj, 745, 149 
\bibitem[Lester et al.(1990)]{Lester+90} Lester, D.~F., Carr, J.~S., Joy, M., \& Gaffney, N.\ 1990, \apj, 352, 544 
\bibitem[Lin et al.(2007)]{Lin+07} Lin, L., Koo, D.~C., Weiner, B.~J., et al.\ 2007, \apjl, 660, L51 
\bibitem[Liu et al.(2018)]{Liu+18} Liu, D., Daddi, E., Dickinson, M., et al.\ 2018, \apj, 853, 172 (L18)
\bibitem[L{\'o}pez-Sanjuan et al.(2010)]{LopezSanjuan+10} L{\'o}pez-Sanjuan, C., Balcells, M., P{\'e}rez-Gonz{\'a}lez, P.~G., et al.\ 2010, \aap, 518, A20 
\bibitem[Lotz et al.(2004)]{Lotz+04} Lotz, J.~M., Primack, J., \& Madau, P.\ 2004, \aj, 128, 163  
\bibitem[Lotz et al.(2008)]{Lotz+08} Lotz, J.~M., Davis, M., Faber, S.~M., et al.\ 2008, \apj, 672, 177
\bibitem[Lotz et al.(2010a)]{Lotz+10a} Lotz, J.~M., Jonsson, P., Cox, T.~J., \& Primack, J.~R.\ 2010, \mnras, 404, 575 
\bibitem[Lotz et al.(2010b)]{Lotz+10} Lotz, J.~M., Jonsson, P., Cox, T.~J., \& Primack, J.~R.\ 2010, \mnras, 404, 590 
\bibitem[Lotz et al.(2011)]{Lotz+11} Lotz, J.~M., Jonsson, P., Cox, T.~J., et al.\ 2011, \apj, 742, 103 
\bibitem[Lutz et al.(2011)]{Lutz+11} Lutz, D., Poglitsch, A., Altieri, B., et al.\ 2011, \aap, 532, A90 
\bibitem[Magdis et al.(2012)]{Magdis+12} Magdis, G.~E., Daddi, E., B{\'e}thermin, M., et al.\ 2012, \apj, 760, 6 
\bibitem[Magnelli et al.(2009)]{Magnelli_2009} Magnelli, B., Elbaz, D., Chary, R.~R., et al.\ 2009, \aap, 496, 57 
\bibitem[Man et al.(2016)]{Man+16} Man, A.~W.~S., Zirm, A.~W., \& Toft, S.\ 2016, \apj, 830, 89 
\bibitem[Mancuso et al.(2016)]{Mancuso+16} Mancuso, C., Lapi, A., Shi, J., et al.\ 2016, \apj, 833, 152 
\bibitem[Mantha et al.(2018)]{Mantha+18} Mantha, K.~B., McIntosh, D.~H., Brennan, R., et al.\ 2018, \mnras, 475, 1549 
\bibitem[Martin et al.(2017)]{Martin+17} Martin, G., Kaviraj, S., Devriendt, J.~E.~G., et al.\ 2017, \mnras, 472, L50 
\bibitem[Meurer et al.(1999)]{Meurer+99} Meurer, G.~R., Heckman, T.~M., \& Calzetti, D.\ 1999, \apj, 521, 64 
\bibitem[Mihos \& Hernquist(1994)]{Mihos_Hernquist_94} Mihos, J.~C., \& Hernquist, L.\ 1994, \apjl, 431, L9 
\bibitem[Mihos \& Hernquist(1994b)]{Mihos_Hernquist_94b} Mihos, J.~C., \& Hernquist, L.\ 1994, \apjl, 425, L13 
\bibitem[Mihos \& Hernquist(1996)]{Mihos_Hernquist_96} Mihos, J.~C., \& Hernquist, L.\ 1996, \apj, 464, 641 
\bibitem[Momcheva et al.(2016)]{Momcheva+16} Momcheva, I.~G., Brammer, G.~B., van Dokkum, P.~G., et al.\ 2016, \apjs, 225, 27 
\bibitem[Moreno et al.(2013)]{Moreno+13} Moreno, J., Bluck, A.~F.~L., Ellison, S.~L., et al.\ 2013, \mnras, 436, 1765 
\bibitem[Mullaney et al.(2011)]{Mullaney+11} Mullaney, J.~R., Alexander, D.~M., Goulding, A.~D., \& Hickox, R.~C.\ 2011, \mnras, 414, 1082 
\bibitem[Murata et al.(2014)]{Murata+14} Murata, K.~L., Kajisawa, M., Taniguchi, Y., et al.\ 2014, \apj, 786, 15 
\bibitem[Murphy et al.(1996)]{Murphy+1996} Murphy, T.~W., Jr., Armus, L., Matthews, K., et al.\ 1996, \aj, 111, 1025 
\bibitem[Naab et al.(1999)]{Naab+99} Naab, T., Burkert, A., \& Hernquist, L.\ 1999, \apjl, 523, L133 
\bibitem[Newman et al.(2012)]{Newman+12} Newman, A.~B., Ellis, R.~S., Bundy, K., \& Treu, T.\ 2012, \apj, 746, 162 
\bibitem[Noeske et al.(2007)]{Noeske+07} Noeske, K.~G., Weiner, B.~J., Faber, S.~M., et al.\ 2007, \apjl, 660, L43 
\bibitem[Nordon et al.(2013)]{Nordon+13} Nordon, R., Lutz, D., Saintonge, A., et al.\ 2013, \apj, 762, 125 
\bibitem[Oesch et al.(2018)]{Oesch+18} Oesch, P.~A., Montes, M., Reddy, N., et al.\ 2018, \apjs, 237, 12 
\bibitem[Oke(1974)]{Oke_1974} Oke, J.~B.\ 1974, \apjs, 27, 21 
\bibitem[Overzier et al.(2011)]{Overzier+11} Overzier, R.~A., Heckman, T.~M., Wang, J., et al.\ 2011, \apjl, 726, L7 
\bibitem[Owen(2018)]{Owen18} Owen, F.~N.\ 2018, \apjs, 235, 34 
\bibitem[Pannella et al.(2009)]{Pannella+09} Pannella, M., Carilli, C.~L., Daddi, E., et al.\ 2009, \apjl, 698, L116 
\bibitem[Pannella et al.(2015)]{Pannella+15} Pannella, M., Elbaz, D., Daddi, E., et al.\ 2015, \apj, 807, 14
\bibitem[Papovich et al.(2005)]{Papovich+05} Papovich, C., Dickinson, M., Giavalisco, M., Conselice, C.~J., \& Ferguson, H.~C.\ 2005, \apj, 631, 101 
\bibitem[Patton et al.(2000)]{Patton+2000} Patton, D.~R., Carlberg, R.~G., Marzke, R.~O., et al.\ 2000, \apj, 536, 153 
\bibitem[Perret et al.(2014)]{Perret+14} Perret, V., Renaud, F., Epinat, B., et al.\ 2014, \aap, 562, A1
\bibitem[Pipino et al.(2014)]{Pipino+14} Pipino, A., Cibinel, A., Tacchella, S., et al.\ 2014, \apj, 797, 127 
\bibitem[Planck Collaboration et al.(2016)]{Planck+16} Planck Collaboration, Ade, P.~A.~R., Aghanim, N., et al.\ 2016, \aap, 594, A13 
\bibitem[Puech et al.(2014)]{Puech+14} Puech, M., Hammer, F., Rodrigues, M., et al.\ 2014, \mnras, 443, L49 
\bibitem[Puglisi et al.(2017)]{Puglisi+17} Puglisi, A., Daddi, E., Renzini, A., et al.\ 2017, \apjl, 838, L18 
\bibitem[Robaina et al.(2009)]{Robaina+09} Robaina, A.~R., Bell, E.~F., Skelton, R.~E., et al.\ 2009, \apj, 704, 324 
\bibitem[Rodighiero et al.(2011)]{Rodighiero+11} Rodighiero, G., Daddi, E., Baronchelli, I., et al.\ 2011, \apjl, 739, L40 
\bibitem[Ryan et al.(2008)]{Ryan+08} Ryan, R.~E., Jr., Cohen, S.~H., Windhorst, R.~A., \& Silk, J.\ 2008, \apj, 678, 751 
\bibitem[Saintonge et al.(2011)]{Saintonge+11} Saintonge, A., Kauffmann, G., Wang, J., et al.\ 2011, \mnras, 415, 61  
\bibitem[Salim et al.(2007)]{Salim+07} Salim, S., Rich, R.~M., Charlot, S., et al.\ 2007, \apjs, 173, 267 
\bibitem[Sanders et al.(1988)]{Sanders+88} Sanders, D.~B., Soifer, B.~T., Elias, J.~H., et al.\ 1988, \apj, 325, 74 
\bibitem[Sanders \& Mirabel(1996)]{Sanders_Mirabel_1996} Sanders, D.~B., \& Mirabel, I.~F.\ 1996, \araa, 34, 749
\bibitem[Santini et al.(2009)]{Santini+09} Santini, P., Fontana, A., Grazian, A., et al.\ 2009, \aap, 504, 751 
\bibitem[Sargent et al.(2012)]{Sargent+12} Sargent, M.~T., B{\'e}thermin, M., Daddi, E., \& Elbaz, D.\ 2012, \apjl, 747, L31 
\bibitem[Sargent et al.(2014)]{Sargent+14} Sargent, M.~T., Daddi, E., B{\'e}thermin, M., et al.\ 2014, \apj, 793, 19 
\bibitem[Schinnerer et al.(2016)]{Schinnerer+16} Schinnerer, E., Groves, B., Sargent, M.~T., et al.\ 2016, \apj, 833, 112 
\bibitem[Schreiber et al.(2015)]{Schreiber+15} Schreiber, C., Pannella, M., Elbaz, D., et al.\ 2015, \aap, 575, A74
\bibitem[Scoville et al.(1997)]{Scoville+97} Scoville, N.~Z., Yun, M.~S., \& Bryant, P.~M.\ 1997, \apj, 484, 702 
\bibitem[Scudder et al.(2012)]{Scudder+12} Scudder, J.~M., Ellison, S.~L., \& Mendel, J.~T.\ 2012, \mnras, 423, 2690 
\bibitem[Scudder et al.(2012b)]{Scudder+12b} Scudder, J.~M., Ellison, S.~L., Torrey, P., Patton, D.~R., \& Mendel, J.~T.\ 2012, \mnras, 426, 549 
\bibitem[Skelton et al.(2014)]{Skelton+14} Skelton, R.~E., Whitaker, K.~E., Momcheva, I.~G., et al.\ 2014, \apjs, 214, 24 
\bibitem[Snyder et al.(2017)]{Snyder+17} Snyder, G.~F., Lotz, J.~M., Rodriguez-Gomez, V., et al.\ 2017, \mnras, 468, 207 
\bibitem[Solomon et al.(1997)]{Solomon+97} Solomon, P.~M., Downes, D., Radford, S.~J.~E., \& Barrett, J.~W.\ 1997, \apj, 478, 144 
\bibitem[Solomon \& Vanden Bout(2005)]{Solomon_VandenBout_2005} Solomon, P.~M., \& Vanden Bout, P.~A.\ 2005, \araa, 43, 677
\bibitem[Sparre \& Springel(2016)]{Sparre_Springel_2016} Sparre, M., \& Springel, V.\ 2016, \mnras, 462, 2418  
\bibitem[Speagle et al.(2014)]{Speagle+14} Speagle, J.~S., Steinhardt, C.~L., Capak, P.~L., \& Silverman, J.~D.\ 2014, \apjs, 214, 15 
\bibitem[Springel(2000)]{Springel2000} Springel, V.\ 2000, \mnras, 312, 859 
\bibitem[Springel et al.(2005)]{Springel+05} Springel, V., Di Matteo, T., \& Hernquist, L.\ 2005, \mnras, 361, 776 
\bibitem[Starkenburg et al.(2016)]{Starkenburg+16} Starkenburg, T.~K., Helmi, A., \& Sales, L.~V.\ 2016, \aap, 595, A56 
\bibitem[Steinhardt et al.(2014)]{Steinhardt+14} Steinhardt, C.~L., Speagle, J.~S., Capak, P., et al.\ 2014, \apjl, 791, L25 
\bibitem[Swinbank et al.(2010)]{Swinbank+10} Swinbank, A.~M., Smail, I., Longmore, S., et al.\ 2010, \nat, 464, 733 
\bibitem[Tacchella et al.(2016)]{Tacchella+16} Tacchella, S., Dekel, A., Carollo, C.~M., et al.\ 2016, \mnras, 457, 2790 
\bibitem[Tacconi et al.(2013)]{Tacconi+13} Tacconi, L.~J., Neri, R., Genzel, R., et al.\ 2013, \apj, 768, 74 
\bibitem[Tacconi et al.(2018)]{Tacconi+18} Tacconi, L.~J., Genzel, R., Saintonge, A., et al.\ 2018, \apj, 853, 179 
\bibitem[Tomczak et al.(2016)]{Tomczak+16} Tomczak, A.~R., Quadri, R.~F., Tran, K.-V.~H., et al.\ 2016, \apj, 817, 118 
\bibitem[Toomre \& Toomre(1972)]{Toomre_Toomre_1972} Toomre, A., \& Toomre, J.\ 1972, \apj, 178, 623 
\bibitem[van Dokkum et al.(2011)]{vanDokkum+11} van Dokkum, P.~G., Brammer, G., Fumagalli, M., et al.\ 2011, \apjl, 743, L15 
\bibitem[van Starkenburg et al.(2008)]{vanStarkenburg+08} van Starkenburg, L., van der Werf, P.~P., Franx, M., et al.\ 2008, \aap, 488, 99 
\bibitem[Veilleux et al.(2002)]{Veilleux+02} Veilleux, S., Kim, D.-C., \& Sanders, D.~B.\ 2002, \apjs, 143, 315 
\bibitem[Vogelsberger et al.(2014)]{Vogelsberger+14} Vogelsberger, M., Genel, S., Springel, V., et al.\ 2014, \mnras, 444, 1518 
\bibitem[Whitaker et al.(2012)]{Whitaker+12} Whitaker, K.~E., van Dokkum, P.~G., Brammer, G., \& Franx, M.\ 2012, \apjl, 754, L29 
\bibitem[Whitaker et al.(2014)]{Whitaker+14} Whitaker, K.~E., Franx, M., Leja, J., et al.\ 2014, \apj, 795, 104 
\bibitem[Williams et al.(2011)]{Williams+11} Williams, R.~J., Quadri, R.~F., \& Franx, M.\ 2011, \apjl, 738, L25 
\bibitem[Willett et al.(2015)]{Willet+15} Willett, K.~W., Schawinski, K., Simmons, B.~D., et al.\ 2015, \mnras, 449, 820 
\bibitem[Wisnioski et al.(2015)]{Wisnioski+15} Wisnioski, E., F{\"o}rster Schreiber, N.~M., Wuyts, S., et al.\ 2015, \apj, 799, 209 
\bibitem[Wong et al.(2011)]{Wong+11} Wong, K.~C., Blanton, M.~R., Burles, S.~M., et al.\ 2011, \apj, 728, 119 
\bibitem[Wuyts et al.(2012)]{Wuyts+12} Wuyts, S., F{\"o}rster Schreiber, N.~M., Genzel, R., et al.\ 2012, \apj, 753, 114 
\bibitem[Xu et al.(2012)]{Xu+12} Xu, C.~K., Shupe, D.~L., B{\'e}thermin, M., et al.\ 2012, \apj, 760, 72 
\bibitem[Xue et al.(2016)]{Xue+16} Xue, Y.~Q., Luo, B., Brandt, W.~N., et al.\ 2016, \apjs, 224, 15 
\bibitem[Zamojski et al.(2007)]{Zamojski+07} Zamojski, M.~A., Schiminovich, D., Rich, R.~M., et al.\ 2007, \apjs, 172, 468 
\bibitem[Zibetti et al.(2009)]{Zibetti_et_al_2009} Zibetti, S., Charlot, S., \& Rix, H.-W.\ 2009, \mnras, 400, 1181 
\bibitem[Zolotov et al.(2015)]{Zolotov+15} Zolotov, A., Dekel, A., Mandelker, N., et al.\ 2015, \mnras, 450, 2327 
\end{thebibliography}
\end{document}